\catcode`\^^M=5 \catcode`\%=14

\documentstyle[prb,aps,amsfonts,floats,epsfig]{revtex}

\def\negmed{}

\def\v#1{{\bbox{#1}}}

\def\vk{{\v k}}
\def\vp{{\v p}}
\def\vq{{\v q}}
\def\vz{{\v z}}
\def\vw{{\v w}}
\def\khat{\v{\hat k}}
\def\phat{\v{\hat p}}
\def\qhat{\v{\hat q}}

\def\Eq(#1){Eq.~(\protect\ref{#1})}
\def\Eqs(#1){Eqs.~(\protect\ref{#1})}
\def\Equation(#1){Equation~(\protect\ref{#1})}
\def\Equations(#1){Equations~(\protect\ref{#1})}
\def\hide#1(#2){(\ref{#2})}
\def\eq(#1){\label{#1}}
\def\eqref(#1){\eqnum{\ref{#1}}}

\def\It#1{{\it #1}}

\def\casefr#1#2{\case#1#2}

\def\Abs#1{{\left|#1\right|}}

\def\<#1>{\left<#1\right>}
\def\ordspacing{}

\def\half{{\casefr12}}	

\def\a{\alpha}
\def\b{\beta}

\def\d{\delta} \def\D{\Delta}
\def\e{\epsilon} 
\def\g{\gamma}

\def\l{\lambda} 
\def\n{\eta} 

\def\s{\sigma} 

\def\Th{\Theta}

\def\fr#1#2{\frac{#1}{#2}}
\def\Partial#1#2{\frac{\partial#1}{\partial#2}}
\def\delt{\Partial{}{t}}

\def\Mathop#1{\mathop{\rm #1}}
\def\Re{\Mathop{Re}}
\def\Im{\Mathop{Im}}

\def\vlp{\mathopen{\hbox{\bf (}}}
\def\vrp{\mathclose{\hbox{\bf )}}}

\let\I\II

\let\exP\Exp

\def\eg#1.{e.g.}
\def\ie#1.{i.e.}

\def\goesto{\rightarrow}

\def\Dirac#1{\delta{\left({\ordspacing #1}\right)}}

\def\be{\begin{equation}}
\def\ee{\end{equation}}

\def\bea{\begin{eqnarray}}
\def\eea{\end{eqnarray}}

\def\SD{\negmed\SDtext\negmed}
\def\SDtext{\sum_{\vk+\vp+\vq=\v0}}

\def\nk{\n_\vk}
\def\np{\n_\vp}
\def\nq{\n_\vq}

\def\Mkpq{M_\kpq}
\def\Mpqk{M_{\vp\vq\vk}}
\def\Mqkp{M_{\vq\vk\vp}}
\def\Mkqp{M_{\vk\vq\vp}}
\def\psik{\psi_\vk}
\def\psip{\psi_\vp}
\def\psiq{\psi_\vq}
\def\linop{\left(\delt + \nuk\right)\!}
\def\kpq{{\vk\vp\vq}}
\def\pqk{{\vp\vq\vk}}

\def\nuk{\nu_\vk}

\def\sigk{\sigma_\vk}
\def\sigp{\sigma_\vp}
\def\sigq{\sigma_\vq}

\def\Ck{C_\vk}
\def\Cp{C_\vp}
\def\Cq{C_\vq}

\def\Rk{R_\vk}
\def\Rp{R_\vp}

\def\Nk{N_\vk}

\def\BTh{\bar\Th}
\def\BThkpq{\bar\Th_\kpq}
\def\BThpqk{\bar\Th_\pqk}

\def\tb{{\bar t}}
\def\sk{\Sigma_\vk}
\def\cFk{{\cal\Fk}}
\def\Fk{F_\vk}

\def\Thkpq{\Theta_\kpq}
\def\Thpqk{\Theta_{\pqk}}

\def\nck{\n_{C \vk}}
\def\ncp{\n_{C \vp}}
\def\ncq{\n_{C \vq}}

\newcommand{\eqr}[1]{Eq.~(\ref{#1})}
\newcommand{\infinity}{\infty}
\def\grad{{\bf \nabla}}

\def\mod{{\rm mod}}
\def\muchl{\ll}

\begin{document}

\draft 

\title{
Non-white noise and a multiple-rate Markovian closure theory for
turbulence
}

\author{Gregory W. Hammett}

\address{Plasma Physics Laboratory, Princeton University, P.O. Box 451,
Princeton, NJ 08543 \\ hammett@princeton.edu}

\author{John C. Bowman}

\address{Department of Mathematical Sciences, University of Alberta, Edmonton,
Alberta, Canada T6G~2G1 \\ bowman@math.ualberta.ca}

\date{Submitted to Physics of Fluids: February 11, 2002}

\maketitle

\begin{abstract}


Markovian models of turbulence can be derived from the renormalized
statistical closure equations of the direct-interaction approximation
(DIA).  Various simplifications are often introduced, including an
assumption that the two-time correlation function is proportional to the
renormalized infinitesimal propagator (Green's function), i.e.~the
decorrelation rate for fluctuations is equal to the decay rate for
perturbations.
While this is a rigorous result of the fluctuation--dissipation theorem
for thermal equilibrium, it does not necessarily apply to all
types of turbulence.
Building on previous work on realizable Markovian closures, we explore
a way to allow the decorrelation and decay rates to differ (which in some
cases affords a more accurate treatment of effects such as non-white noise),
while retaining the computational advantages of a Markovian approximation.
Some Markovian approximations differ only in the initial transient phase,
but the multiple-rate Markovian closure (MRMC) presented here could modify
the steady-state spectra as well.
Markovian models can be used directly in studying turbulence in a wide
range of physical problems (including zonal flows, of recent interest in
plasma physics), or they may be a useful starting point for deriving
subgrid turbulence models for computer simulations.

\end{abstract}

\pacs{PACS: 47.27.Eq, 47.27.Sd, 05.40.-a}

%

\twocolumn
\narrowtext


\section{Introduction}\label{intro}

Our derivation builds on and closely follows the work by Bowman, Krommes,
and Ottaviani\cite{Bowman93} (we will frequently refer to this paper as
BKO), on realizable Markovian closures derived from Kraichnan's
direct-interaction-approximation (the DIA).  The DIA is based on a
renormalized perturbation theory and gives an integro-differential set of
equations to determine the two-time correlation function.  The DIA involves
time integrals over the past history of the system, which can be
computationally expensive.  Markovian approximations give a simpler set of
differential equations that involve only information from the present time.
They approximate two-time information in the correlation function and in the
renormalized Green's function by a decorrelation rate parameter.  The
structure of the equations we derive here is similar to the realizable
Markovian closure (RMC) of BKO,\cite{Bowman93} but with extensions such as
replacing a single decorrelation rate parameter with several different
nonlinear rate parameters, to allow for a more accurate model of effects such
as non-white noise.  (As will be discussed more below, the RMC does include 
more non-white-noise effects than one might think at first.)

The basic issue studied in the present paper can be illustrated by a simple
Langevin equation (which will be discussed in more detail in the next
section):
\begin{equation}
\left( {\partial \over \partial t} + \eta \right) \psi(t) = f(t),
\label{Langevin-intro}
\end{equation}
where $\eta$ is the decay rate and $f$ is a random forcing or stirring term
(also known as noise).  As is well known, if $f$ is white noise, then the
decorrelation rate for $\psi$ is given by $\eta$, so that in a statistical
steady state the two-time correlation function $\langle \psi(t) \psi^*(t')
\rangle = C_0 \exp(-\eta |t-t'|)$ (assuming constant real $\eta$ here).
However, if $f(t)$ varies slowly compared to the 1/$\eta$ time scale, then
the solution to the Langevin equation is just $\psi(t) \approx f(t) /\eta$,
and the decorrelation rate for $\psi$ is instead given by the decorrelation
rate for $f$.  Note that the Green's function (the response to a
perturbation at time $t'$) is still $\exp(-\eta (t-t'))$.  Previous
Markovian closures employed some variant of an ansatz, based on the
fluctuation--dissipation theorem, that the two-time correlation function
and the Green's function were proportional to each other.  This is a
rigorous result for a system in thermal equilibrium, but may not
necessarily apply to a turbulent system.  The purpose of the present paper
is to explore an extended Markovian closure, which we will call the
Multiple-Rate Markovian Closure (MRMC), that allows the decorrelation
rate of $\psi$ to differ from the decay rate $\eta$.


In practice, the corrections due to non-white-noise effects may be
quantitatively modest, as the decorrelation rate for the turbulent noise
$f$ that is driving $\psi$ at a particular wave number $\vk$ is often
comparable to or greater than the nonlinear damping rate $\eta$ at that
$\vk$.  This is because the turbulent noise driving mode $\vk$ arises from the
nonlinear beating of other modes $\vp$ and $\vq$ such that $\vk = \vp +
\vq$.  Thus $|\vp|$ or $|\vq|$ has to be comparable to or larger than
$|\vk|$, and will thus have comparable or larger decorrelation rates, since
the decay rate $\eta$ is usually an increasing function of $|\vk|$.
Furthermore, there are some offsetting effects due to the time-history
integrals in the DIA's generalized Langevin equation that might further
reduce the difference between the decay rate and the decorrelation rate.
Indeed, past comparisons of the RMC with the full DIA or with the full
nonlinear dynamics have generally found fairly good agreement in many
cases,\cite{Bowman93,Bowman97,Krommes2002,LoDestro91} including two-field
Hasegawa--Wakatani
drift-wave turbulence\cite{Hu95,Hu97} and galactic dynamo MHD
turbulence.\cite{Chandran97}
%
Some of the results in this paper help to give a deeper insight into why
this agreement is often fairly good, despite the arguments of the
previous paragraph, i.e., why the fluctuation--dissipation
ansatz is often a reasonable approximation even out of thermal
equilibrium.
But there may be some regimes where the differences are important and the
improvements suggested here would be welcome.  These might include 
include plasma cases where the wave dynamics can make $\eta$ vary strongly
with the direction of $\vk$ in some cases (with strong Landau damping in
some directions and strong instabilities in other directions, for
example), or non-steady-state cases involving zonal flows exhibiting
predator-prey dynamics.

Markovian closures such as the test-field model (TFM) or Orszag's
eddy-damped quasinormal Markovian (EDQNM) closure have been extensively
used to study turbulence in incompressible fluids and plasmas.  The
introduction of BKO\cite{Bowman93} provides useful
discussions of the background of the DIA and Markovian closures, and we
will add just a few remarks here (there are also many reviews on these
topics, such as
Refs.~\onlinecite{Krommes2002,Kadomtsev65,Leslie73b,McComb91,Orszag73,Frederiksen94}).
The RMC developed in BKO\cite{Bowman93} is similar to the
EDQNM, but has features that ensure realizability even in the presence of the
linear wave phenomena exhibited by plasmas (e.g. drift waves) and rotating
planetary flows (e.g. Rossby waves).
``Realizability'' is a property of a statistical 
closure approximation that ensures that, even though it is only an
approximate solution of the original equations, it is an exact solution to
some other underlying stochastic equation, such as a Langevin equation.
The absence of
realizability can cause serious physical and numerical problems, such as
the prediction of negative or even divergent energies.  The RMC reduces to
the DIA-based version of the EDQNM in a statistical steady state, so in some
cases the issue of realizability is only important in the transient phase
as a steady state is approached or in freely decaying turbulence.
Realizability may also be important in certain cases of recent interest
among fusion researchers where oscillations may occur between various parts
of the spectrum (such as predator--prey type oscillations between drift
waves and zonal flows\cite{Leboeuf93,ZLin99}) where a simple
statistical steady state might not exist, or where one is interested in the
transient dynamics.  Unlike some Markovian models that differ only in the
transient dynamics, the Multiple-Rate Markovian Closure presented here
could also alter the steady-state spectrum.

Our results apply to a Markovian approximation of the DIA for a generic
one-field system with a quadratic nonlinearity.  They are immediately
applicable to some simple drift-wave plasma turbulence problems,
Rossby-wave problems, or two-dimensional hydrodynamics. Future work
could extend this approach to 
multiple fields, similar to the covariant multifield RMC of
BKO\cite{Bowman93} or their later realizable test-field
model.\cite{Bowman97}  Multiple field equations can get computationally
difficult (with the compute time scaling as $n^6$, where $n$ is the
number of fields), though two-field studies have
been done\cite{Hu97} and 
disparate scale approximations\cite{Krommes2000b} or other
approximations\cite{Ottaviani91} might make them more tractable.
In addition to their direct use in studying turbulence in a wide
range of systems, the Markovian closures discussed here might also be
useful in deriving subgrid turbulence models for computer
simulations.\cite{Kraichnan76,Frederiksen99}

While our formulation is general and potentially applicable to a wide range
of nonlinear problems involving Markovian approximations, we were motivated
by some recent problems of interest in plasma physics and fusion energy
research, such as zonal
flows.\cite{Hasegawa87,Carreras91,Diamond91,Hammett93}
Initial analytic work elucidating the essentials of nonlinear zonal flow
generation used weak-turbulence
approximations\cite{Diamond98,Smolyakov2000} or
secondary-instability analysis.\cite{Drake92}  Recent interesting work by
Krommes and Kim\cite{Krommes2000b} uses a Markovian statistical theory to
extend the study of zonal flows to the strong turbulence regime.  An
important question is why the strong generation of zonal flows seen near
marginal stability is not as important in stronger instability regimes
(i.e., why is the Dimits nonlinear shift
finite?).\cite{Dimits00,Rogers2000,Dorland2000}
A strong turbulence theory is needed to study this.  An alternative
approach,\cite{Rogers2000,Dorland2000} which has been fruitful in providing
the main
answers to the finite Dimits shift question, is to analyze the secondary
and tertiary instabilities involved in the generation and breakup of zonal
flows.  That work suggests that a complete strong-turbulence Markovian
model of this problem would also need multi-field and geometrical
effects (involving at least the potential and temperature fields, along
with certain neoclassical effects in toroidal magnetic field geometry).  
%
%



Based on the reasoning immediately following \eqr{Langevin-intro} above, 
one might think that the assumption that
the two-time correlation function and the Green's function are proportional
to each other is rigorous only in the limit of white noise (which has an
infinite
decorrelation rate).  The Realizable Markovian Closure has been shown to
correspond exactly to a simple Langevin equation (where the effects of the
turbulence appear in nonlinear damping and nonlinear noise terms), for
which this might appear to be the implication.  However, the mapping from
statistically averaged equations (such as Markovian closures) back to a
stochastic equation for which it is the solution, is not necessarily
unique.  In particular, the full DIA corresponds to a \It{generalized
Langevin} equation (\Eq(generalized-Langevin) below), in which the
damping term $\eta \psi(t)$ in the simple
Langevin equation is replaced by a time-history integral operator.  As
we will find, it is
then possible for the two-time correlation function and the Green's
function to be proportional to each other even when the noise has a finite
correlation time.  This allows the fluctuation--dissipation theorem (which is
rigorous in thermal equilibrium) to be satisfied without requiring the
noise to be white (since the noise is not necessarily white in thermal
equilibrium).  
%
%
%
%
Thus the fluctuation--dissipation ansatz of BKO is a less restrictive
assumption than one might have at first thought.  [It should be noted that
previous Markovian models account implicitly for at least some
non-white-noise effects.  For example, in the calculation of the triad
interaction time $\theta_{\vk \vp \vq} = 1 / (\nk + \np +\nq)$ for
three-wave interactions, finite values of the assumed noise
decorrelation rate $\np + \nq$ are used.]

Nevertheless, there is still no reason in a general situation that the
two-time correlation function and the Green's function be constrained to
be proportional to each other.  As described elsewhere, there may be
regimes where the resulting differences between the decorrelation rate
and the decay rate are significant.

%
%


The outline of this paper is as follows.  Sec.~(\ref{sec-simple}) presents
some of the essential ideas of this paper for a very simple Langevin
equation, and includes a section motivating the choice of the limit
operator introduced by BKO\cite{Bowman93} to ensure realizability.
Sec.~(\ref{steady-state-Langevin}) presents a more detailed calculation of
the non-white Markovian model for a simple Langevin equation, including the
effects of complex damping rates (to represent the wave frequency) and the
issue of Galilean invariance.  The model is compared with exact results in
the steady-state limit, and then an extension to time-dependent Langevin
statistics is presented (along with, in Appendix~\ref{Appendix-real}, an
alternative proof of realizability for this case). 
Sec.~(\ref{closures}) presents the notation of
the full many-mode nonlinear equations we will solve and summarizes the
direct interaction approximation (DIA), which is our starting point.
Sec.~(\ref{sec-Markov-ss}) summarizes how the non-white Markovian
approach is derived for the steady-state limit (with further details given
in Appendix~\ref{Appendix-fit}), while
Sec.~(\ref{sec-Markov}) presents the full non-white Markovian approximation
for the time-dependent DIA.
Sec.~(\ref{Markov-properties}) discusses some important properties of
these equations, including the limits of thermal equilibrium and
inertial range scaling, and some difficulties due to the lack of random
Galilean invariance in the DIA and described in
Appendix~\ref{Appendix-scaling}.
The conclusions include some suggestions for future research.


\section{Simple examples based on the Langevin equation}
\label{sec-simple}

Here we expand upon the analogy given in Sec.~(\ref{intro}) using a
simple Langevin equation, which provides a useful paradigm for understanding
the essential ideas considered in this paper.  Since realizable Markovian
closure approximations to the DIA can be shown to correspond exactly to an
underlying set of coupled Langevin equations, the analogy is quite
relevant.  In this section we will consider heuristic arguments based on
some simple scalings; later sections will be more rigorous.


\label{intro-non-white}


Consider the simple Langevin equation
\begin{equation}
\left( {\partial \over \partial t}  + \eta(t) \right) \psi = f^*(t),
\label{Langevin}
\end{equation}
where $\eta$ is a damping rate and $f^*$ is a random forcing or stirring
term (also known as ``noise'').  [Here we now force
\eqr{Langevin-intro} with the complex conjugate of $f$, for
consistency with the form of the equations used later for a generic
quadratically nonlinear equation.]  The statistics of the noise are given
by a specified
two-time correlation function $C_f(t,t') = \langle f(t) f^*(t') \rangle$.  [In
the white-noise limit, $C_f(t,t') = 2 D \delta(t-t')$, and the power
spectrum of the Fourier-transform of $f(t)$ is independent of frequency,
and is thus called a ``white'' spectrum.]  The Langevin equation is used to
model many kinds of systems exhibiting random-walk or Brownian motion
features.  Here we can think of $\psi$ as the complex amplitude of one
component of the turbulence with a specified Fourier wave number $\vk$.
Note that $\eta$ may be complex (representing both damping and wave-like motions) and
represents both linear and nonlinear (renormalized) damping or frequency
shifts due to interactions with other modes.  The random forcing $f^*$
represents nonlinear driving by other modes beating together to drive this
mode.

The response function (or Green's function or propagator) for this equation
satisfies
\begin{equation}
\left( {\partial \over \partial t } + \eta \right) R(t,t') = \delta(t-t'),
\label{Langevin-Green}
\end{equation}
which easily yields $R(t,t')=\exp(-\eta (t-t')) H(t-t')$ if $\eta$ is
independent of time, where $H(t)$ is
the Heaviside step function.  The solution to the Langevin equation is just
$\psi(t) = \int_0^t d \bar{t} \, R(t,\bar{t}) f^*(\bar{t})$ (for the initial
condition $\psi(0)=0$).  It is then straightforward to demonstrate the
standard result that, if $f$ is white noise and the long-time statistical
steady-state limit is considered, then the correlation function for $\psi$
is
\[
C(t,t') \doteq \langle \psi(t) \psi^*(t') \rangle = C_0
\exp(-\eta(t-t'))
\]
for $t>t'$, where $C_0 = 2 D / (\eta + \eta^*)$ (we emphasize definitions
with the notation $\doteq$).  [For $t<t'$, one can use
the symmetry condition $C(t,t') = C^*(t',t)$.]  That is, the decorrelation rate
for $\psi$ is just $\eta$.  This is equivalent to the assumption in a broad
class of Markovian models that the decorrelation rate for $\psi$ is the
same as the decay rate of the response function.

However, consider the opposite of the white-noise limit, where $f$ varies
slowly in time compared to the $1/\eta$ time scale.  Then the solution of
\eqr{Langevin} is approximately $\psi(t) = f^*(t)/\eta$, and the
decorrelation rate for $\psi$ will be the same as the decorrelation rate
for $f^*$.  In this limit, the assumption in many Markovian models that the
decorrelation rate is $\eta$ is not valid.

Denote the decorrelation rate for $f^*$ as $\eta_f^*$, and the decorrelation
rate for $\psi$ as $\eta_C$.  Then one might guess that a simple
Pad\'e-type formula that roughly interpolates between the white-noise limit
$\eta_f \gg \eta$ and the opposite ``red-noise'' limit $\eta_f \ll \eta$
would be something like $1/\eta_C \approx 1/\eta + 1/\eta_f^*$, or
\begin{equation}
\eta_C = {\eta \, \eta_f^* \over \eta + \eta_f^*}.
\label{Intro-Pade}
\end{equation}
In fact, we will discover in the next section that more detailed
calculations give similar results in the limit of real $\eta$ and $\eta_f$,
though the formulas are more complicated in the presence of wave behavior
with complex $\eta$ and $\eta_f$.

We note that in many cases of interest, the noise decorrelation rate
$\eta_f$ turns out to be of comparable magnitude to $\eta$ (for example,
if the dominant interactions involve modes of comparable scale).  In
this case, while the white-noise approximation is not rigorously valid,
the corrections to the decorrelation rate considered in this paper might
turn out to be quantitatively modest, $\sim 50\%$.  Furthermore, in the
case of the full DIA and its corresponding generalized Langevin
equation, we will find additional corrections that can, in some cases,
offset the effects in \eqr{Intro-Pade} and cause $\eta_C$ to be closer
to $\eta$.



Before going on to the more detailed results in the next section, we
consider the meaning of an operator introduced in the BKO\cite{Bowman93}
derivations in order to preserve realizability in the time-dependent case,
where $\eta(t)$ varies in time and may be negative (transiently),
representing an instability.  [In order for a meaningful long-time
steady-state limit to exist, the net $\eta$ (which is the sum of linear and
nonlinear terms) must eventually go positive to provide a sink for the
noise term.  But it is important to preserve realizability during the
transient times when $\eta$ may be negative.]  Based on arguments about
symmetry and the steady-state fluctuation--dissipation theorem, they
initially proposed a time-dependent ansatz of the form
\begin{equation}
C(t,t') = C^{1/2}(t) C^{1/2}(t') \exp \left(-\int_{t'}^t d \bar{t} \, \,
      \eta(\bar{t}) \right)
\label{FDansatz}
\end{equation}
(for $t>t'$), where $C(t) \doteq C(t,t)$ is the equal-time covariance.
Later in their derivation, they state that in order to ensure
realizability, $\eta(\bar{t})$ in this expression had to be replaced with
${\cal P}(\eta(\bar{t}))$, where the operator ${\cal P}(\eta) = \Re \eta
H(\Re \eta) + i \Im \eta$ prevents the real part of the effective $\eta$
in \eqr{FDansatz} from going negative.

Physically this makes sense for the following reasons.  Consider
\eqr{Langevin} with white noise $f$ (thus ignoring the non-white-noise
effects).  Then in a normal statistical steady state where $\eta(t)$ is
constant and $\Re \eta > 0$, \eqr{FDansatz} properly reproduces the usual
result $C(t,t')= C_0 \exp(-\eta|t-t'|)$.  However, if $\Re \eta<0$ (which
it might do at least transiently in the full turbulent system considered
later), then \eqr{Langevin} can't reach a steady state, and the solution is
eventually $\psi(t) = \psi_0 \exp(-\eta t) = |\psi(t)| \exp(- i \Im \eta
t)$, after an initial transient phase.    Thus
$C(t,t')=C^{1/2}(t) C^{1/2}(t') \exp(-i \Im \eta(t-t'))$, in agreement
with and providing an additional intuitive argument for
BKO's modified form of \eqr{FDansatz}, including the ${\cal P}(\eta)$
operator.
[There may be an initial phase
where the noise term $f$ in \eqr{Langevin} dominates and causes $C(t)$ to
grow linearly in time, $C(t) = \langle \psi(t) \psi^*(t) \rangle = 2 D
t$. But eventually the unstable $\eta \psi$ term will become large enough
to dominate and lead to exponential growth of $\psi$.]


The model we will introduce below replaces $\eta$ in \Eq(FDansatz) with a
separate parameter $\eta_C$, and develops a formula to relate $\eta_C$ to
other parameters in the problem such as $\eta$ and $\eta_f$.  In the
white-noise limit, the formula for $\eta_C$ automatically reproduces the
effects of the ${\cal P}$ limiting operator, as will be described in the
next section and in Appendix~(\ref{Appendix-real}).  But numerical
investigation of non-white noise with wave dynamics ($\Im \eta \neq 0$ or
$\Im \eta_f \neq 0$) uncovered cases where the ${\cal P}$ limiting operator is
still needed to ensure realizability.  This will be explained at the end of
Sec.~(\ref{Sec-Time-dependent-Langevin}).

We considered naming the method described in this paper the Non-White
Markovian Closure since, for the simple Langevin equation considered here
and in the next section, the decorrelation rate and the decay rate are
equal only in the white-noise limit, and this approach allows these rates
to differ.  [Alternatively, to emphasize the flexibility of this method one
might have called it the Colored-Noise Markovian Closure since instead of
being restricted to white-noise (a uniform spectrum) we can allow a noise
spectrum of width $\delta \omega \sim \Re \eta_f$
peaked near an arbitrary frequency $\omega \sim \Im \eta_f$.
In other words, this closure can
model spectra with a range of possible colors.]  However, as we will
discuss further, while a simple Langevin equation is sometimes used to
demonstrate realizability of Markovian approximations, the DIA is actually
based on a generalized Langevin equation involving a non-local time-history
integral (compare \Eq(Langevin) with \Eq(generalized-Langevin)).  Because
non-white fluctuations enter not only by making the noise term non-white
but also by affecting this time-history integral, it is
possible for the decay-rate and the decorrelation rate to be equal even in
some cases where the noise is not white (as indeed is the case in thermal
equilibrium where the fluctuation--dissipation theorem must hold but the
noise is not necessarily white).  We thus
favor the name Multiple-Rate Markovian Closure (MRMC), to emphasize
that the method developed here is a generalization of the previous
Realizable Markovian Closure (RMC) to allow for multiple rates (i.e.,
separate decay and decorrelation rates).




\section{Detailed demonstration of the Multiple-Rate Markovian method
with the Langevin equation}\label{steady-state-Langevin}

In this section, we demonstrate the Multiple-Rate Markovian approach
starting with a simple Langevin equation.  The steps in the
derivation are quite similar to the steps that will be taken in the
following sections for the case 
of the more complete DIA for more complicated
nonlinear problems, and thus help build insight and familiarity.  In this
section, we will be introducing various approximations that may seem
unnecessary for the simple Langevin problem, which can be solved exactly in
many cases (for simple forms of the noise correlation function).  But
these are the same approximations that will be used later in deriving
Markovian approximations to the DIA, and so it is useful to be able to test
their accuracy in the Langevin case.

Our starting point is the Langevin \Eq(Langevin), but we
allow $\eta(t)$ to be a function of time, so that the solution to
\Eq(Langevin-Green) for the response function is
\begin{equation}
R(t,t') = \exp\left(-\int_{t'}^t d\bar{t} \, \eta(\bar{t})\right) H(t-t')
\label{Langevin-R}
\end{equation}
(instead of the solution given immediately after \Eq(Langevin-Green), which
assumes that $\eta$ is independent of time).  The solution to the Langevin
equation is
\begin{equation}
\psi(t) = R(t,0) \psi(0) + \int_0^t d \bar{t}\, R(t,\bar{t}) f^*(\bar{t}).
\label{Langevin-solution}
\end{equation}
In principle it is possible to calculate directly two-time statistics like
$C(t,t') = \langle \psi(t) \psi^*(t') \rangle$ from this, but in practice it
is often convenient to consider instead the differential equation for
$\partial C(t,t') / \partial t$, which from \Eq(Langevin) and
\Eq(Langevin-solution) is
\begin{eqnarray}
\left( {\partial \over \partial t} + \eta \right) C(t,t') & = & 
\langle f^*(t) \psi^*(t') \rangle  \nonumber \\
 & = & \int_0^{t'} d \bar{t} \, R^*(t',\bar{t}) C_f^*(t, \bar{t}),
\label{Langevin-2time}
\end{eqnarray}
where the noise correlation function is defined as $C_f(t,t') = \langle
(f(t) f^*(t') \rangle$, and we have assumed that the initial condition $\psi(0)$
has a random phase.  This equation is the analog of the DIA equations
for the two-time correlation function (compare with \Eq(closure a) and
\Eqs(DIA)), but 
with an integral only over the noise and no nonlinear modification of the
damping term.

We define the \It{equal-time correlation function} $C(t)$ in terms of the
\It{two-time correlation function} $C(t,t')$ as $C(t) = C(t,t) =\langle
\psi(t) \psi^*(t') \rangle$  (note that these two functions are
distinguished only by the number of arguments).  Then
\begin{equation}
{\partial C(t) \over \partial t} + 2 \Re \eta \, C(t)
= 2 \Re \int_0^{t} d \bar{t}\, R^*(t,\bar{t}) C_f^*(t,\bar{t}).
\label{Langevin-1time}
\end{equation}
This is the analog of the DIA equal-time covariance equation, \Eq(DIACeqBTh).

\subsection{Langevin statistics in the steady-state limit}
\label{Sec-Langevin-ss}

Consider the steady-state limit where $t, t' \rightarrow \infinity$ (but
with finite time separation $t-t'$), and assume the noise correlation
function has the simple form $C_f(t,t') = C_{f 0} \exp[-\eta_f (t-t')]$ for $t>t'$.  In this
section we assume $\eta$ and $\eta_f$ are time-independent constants.
The response
function reduces back to its steady-state form $R(t,t') = \exp[-\eta(t-t')]
H(t-t')$.  Then \Eq(Langevin-1time) in steady state gives 
\begin{equation}
C_0 \doteq
\lim_{t \rightarrow \infinity} C(t) =C_{f0} \fr{\Re(\eta+\eta_f)}{\Re(\eta)
(\eta +\eta_f) (\eta^* + \eta_f^*)}.
\end{equation}
Writing $\eta = \nu + i \omega$ and
$\eta_f = \nu_f + i \omega_f$ in terms of their real and imaginary
components, and denoting the frequency mismatch $\Delta \omega = \omega +
\omega_f$ (remember, because the complex conjugate $f^*$ is used as the
forcing term, resonance occurs when $\Im(\eta) = \Im(\eta_f^*)$)
this can be written as
\begin{equation}
C_0 = {C_{f0} \over \nu} {(\nu + \nu_f) \over (\nu + \nu_f)^2 + (\Delta
\omega)^2 }.
\label{Langevin-C0}
\end{equation}
This has a familiar Lorentzian form characteristic of resonances.

To find the two-time correlation function, the time integral in
\Eq(Langevin-2time) can be evaluated for $t>t'$ to give
\begin{equation}
\left( {\partial \over \partial t} + \eta \right) C(t,t')
 = { C_{f 0} \over \eta^* + \eta_f^* } \exp[-\eta_f^*(t-t')].
\label{Langevin-ss-2time}
\end{equation}
With the steady-state boundary condition $C(t=t',t') = C_0$, this can be
solved to give 
\begin{eqnarray}
C(t,t') & = & C_0 \left[ 1 - {\Re(\eta) (\eta + \eta_f) \over \Re(\eta +
 \eta_f) (\eta - \eta_f^*) } \right] \exp[-\eta(t-t')] \nonumber \\
 & + & C_0 {\Re(\eta) (\eta+\eta_f) \over \Re(\eta+\eta_f) (\eta-\eta_f^*)}
\exp[-\eta_f^* (t-t')].
\label{Langevin-ss-C2}
\end{eqnarray}
In the white-noise limit, $|\eta_f| \gg |\eta|$, this reduces to the
standard simple result $C(t,t')=C_0 \exp[-\eta(t-t')]$.  But in the more
general case of non-white noise, the two-time correlation function is more
complicated.  [Despite the apparent singularity in the denominator, it is
cancelled by the exponentials so that $C(t,t')$ is well-behaved in the
limit $\eta \rightarrow \eta_f^*$.]  In the context of the turbulent
interaction of many modes, $C_f(t,t')$ and thus $C(t,t')$ may be very
complicated functions.  Even if the noise correlation function has a simple
exponential dependence $C_f(t,t') \propto \exp[-\eta_f(t-t')]$, we see that
the resulting correlation function for $\psi$ is more complicated.  

Consider the task of fitting this complicated $C(t,t')$ with a simpler
model of the form
\be
C_{\mod}(t,t') = C_0 \exp[-\eta_C (t-t')]
\label{Cmod-ss}
\ee
(for $t>t'$).  One way to define the effective decorrelation rate
$\eta_C$ might be based on the area under the time integral,
\begin{equation}
\int_{-\infinity'}^t dt' \, C_{\mod}(t,t') =
{C_0 \over \eta_C} = \int_{-\infinity'}^t dt' \, C(t,t').
\label{etac-2crude}
\end{equation}
This can be evaluated either by directly substituting \Eq(Langevin-ss-C2),
or by taking a time average of \Eq(Langevin-ss-2time); the same answer
results either way. It turns out that in the later versions of this
calculation it is easier to determine $\eta_C$ by integrating the
governing differential equation over time.  Operating on
\Eq(Langevin-ss-2time) with $\int_{-\infinity}^t dt'$ and using
\begin{equation}
\int_{-\infinity}^t dt' \, {\partial C(t,t') \over \partial t} = 
{\partial \over \partial t} \int_{-\infinity}^t dt' \, C(t,t') \, - C(t,t),
\label{Langevin-swap-t}
\end{equation}
we find
\begin{equation}
{1 \over \eta_C} = {1 \over \eta} + {\Re(\eta) (\eta + \eta_f) \over
\Re(\eta + \eta_f) \, \eta \, \eta_f^* }.
\label{etac-2pole}
\end{equation}
This recovers the white-noise limit $\eta_f \gg \eta$ and the red-noise
limit $\eta_f \ll \eta$ discussed in Sec.~\ref{intro-non-white}.  In the
limit of real $\eta$ and real $\eta_f$
it simplifies to the Pad\'e approximation $\eta_C = \eta \eta_f/(\eta +
\eta_f)$ also suggested in the introduction.  However, there is a problem
with \Eq(etac-2pole) related to Galilean invariance.  Suppose we make the
substitutions $\psi = \hat{\psi} \exp[i \omega_2 t]$ and $f^* = \hat{f}^*
\exp[i \omega_2 t]$ into the Langevin \Eq(Langevin).  Then it can
be written as
\begin{equation}
\left( {\partial \over \partial t}  + \hat{\eta} \right) \hat{\psi} =
\hat{f}^*(t),
\end{equation}
where $\hat{\eta} = \eta + i \omega_2$, and the results should be the
same if written in terms of the transformed variables.  In particular, the
correlation function should transform as $\langle \hat{\psi}(t)
\hat{\psi}^*(t') \rangle =\exp[-i \omega_2 (t-t')] \langle \psi(t) \psi^*(t')
\rangle =\exp[-i \omega_2 (t-t')] C(t,t')$.  Thus the decorrelation rate
$\hat{\eta}_C$ for $\hat{\psi}$ should be related to the decorrelation rate
$\eta_C$ for $\psi$ by $\hat{\eta}_C = \eta_C + i \omega_2$. 
The decorrelation rate for the transformed noise term $\hat{f}^*$ also
transforms as $\hat{\eta}_f^* = \eta_f^* + i \omega_2$.  In the case of fluid
or plasma turbulence where $\psi$ represents the amplitude of a Fourier
mode $\propto \exp[i \vk \cdot {\v x}]$ and $f^*$ represents the amplitude
of two modes with wave numbers $\vp$ and $\vq$ beating together to drive
the $\vk$ mode (so $\vp + \vq = \vk$), these transformations correspond to
a Galilean transformation to a moving frame $ {\v x} = {\v x}_0
+ {\v v} t $, with $\omega_2 = \vk \cdot {\v v}$.

So all results should be independent of $\omega_2$ under the transformation
$\eta = \hat{\eta} - i \omega_2$, $\eta_f^* = \hat{\eta}_f^* - i \omega_2$,
(thus $\eta_f = \hat{\eta}_f + i \omega_2$), $\eta_C = \hat{\eta}_C - i
\omega_2$.  \Eq(Langevin-C0) satisfies this, but \Eq(etac-2pole) fails this
test.  This problem and its solution is described in the review paper by
Krommes,\cite{Krommes2000-invariance} who shows it is related to other
differences in various previous Markovian closures.  The problem can be
traced to the definition of \Eq(etac-2crude), which doesn't satisfy the
invariance for general forms of $C(t,t')$.  For example, we could have
multiplied the integrand in \Eq(etac-2crude) by an arbitrary weight
function (such as $\exp[- i \omega_2 (t-t')]$) before taking the time average,
and the results would have changed. The way to fix this problem is to do
the time average in a natural frame of reference for $\psi$ that accounts
for its frequency dependence.  This leads us to the
definition:
\begin{equation}
{C_0^2 \over \eta_C + \eta_C^*} \doteq \int_{-\infinity}^t dt' \,
C_{\mod}^*(t,t') C(t,t').
\label{etac-invariant}
\end{equation}
This corresponds to fitting $C_{\mod}(t,t')$ to $C(t,t')$ by requiring that
both effectively have the same projection onto the function
$C_{\mod}(t,t')$.  [As Krommes\cite{Krommes2000-invariance} points out,
using the invariant definition \Eq(etac-invariant) instead of
\Eq(etac-2crude) is a non-trivial point needed to ensure
realizability and avoid spurious nonphysical solutions in some cases.]

Operating on \Eq(Langevin-ss-2time) with $\int_{-\infinity}^t dt' \,
C_{\mod}^*(t,t')$, using a generalization of \Eq(Langevin-swap-t), and doing
a little rearranging yields
\begin{equation}
\eta_C = \eta - {C_{f0} (\eta_C +\eta_C^*) \over C_0 (\eta^* + \eta_f^*)
(\eta_C^* + \eta_f^*) }.
\label{Langevin-eta_c-ss}
\end{equation}
This is properly invariant to the transformation described in the previous
paragraph. Solving for $\eta_C$ while leaving $\eta_C^*$ on the other side of
the equation, eventually leads to
\begin{equation}
\eta_C = { \eta \eta_f^* \Re(\eta + \eta_f) + i \eta_C^* \Im(\eta \eta_f^*)
\over
(\eta + \eta_C^*) \Re(\eta + \eta_f) + (\eta_f^* + \eta^*) \Re(\eta_f) }.
\label{eta_c-ss}
\end{equation}
If we consider the limit where $\eta$, $\eta_f$, and thus $\eta_C$ are all
real, this simplifies to the form
\begin{equation}
\eta_C = {\eta \eta_f \over \eta + \eta_f + \eta_C}.
\label{eta_c-real}
\end{equation}
This is similar to (but more accurate than) the rough interpolation formula
\Eq(Intro-Pade) suggested in the introduction.  This kind of recursive
definition, with $\eta_C$ appearing on both sides, is a common feature of
the steady-state limit of theories based on the renormalized DIA equations,
and can be solved in practice by iteration, or by considering the
time-dependent versions of the theories.  In \Eq(eta_c-real) with real
coefficients, one can easily solve this equation for $\eta_C$, but the
solution is much more difficult in the case of complex coefficients in
\Eq(eta_c-ss).
The resulting calculation is laborious, so we used the symbolic algebra
package Maple\cite{Maple} to solve for $\eta_C$ with complex
coefficients.  Looking at the real and imaginary parts of \Eq(eta_c-ss)
separately eventually leads to a quadratic equation and a linear equation
to determine the real and imaginary parts of $\eta_C$.  Unfortunately it
takes 16 lines of code to write down the resulting closed-form solution
(though perhaps there are common subexpressions that would simplify it).
(Maple worksheets that show this calculation and check other main results
in this paper are available online.\cite{Maple-greg-results})
%
%
This is tedious for humans but easy to evaluate in Fortran, C, or other
computer language.  On the other hand, this is only helpful for the simple
Langevin problem anyway since direct solution is not really practical for
the full nonlinear problem considered by the DIA, where the noise term of
the Langevin equation is replaced by a sum over many modes.  In many cases
of interest, the noise decorrelation rate $\eta_f$ turns out to be
comparable in magnitude to $\eta$, so iteration of \Eq(eta_c-ss) usually
converges quickly.  
(However, there are limits where convergence is slow, such as some strongly
non-resonant cases where $\Re \eta$ is very close to $\Re \eta_f$ and both
are very small compared to $\Im(\eta_f+\eta)$.)  The other option is to
consider the time-dependent problem, the topic of the subsection after
next, which effectively performs an iteration in time as a steady state
is approached.

\begin{figure}[!ht]
\begin{center}
\epsfig{file=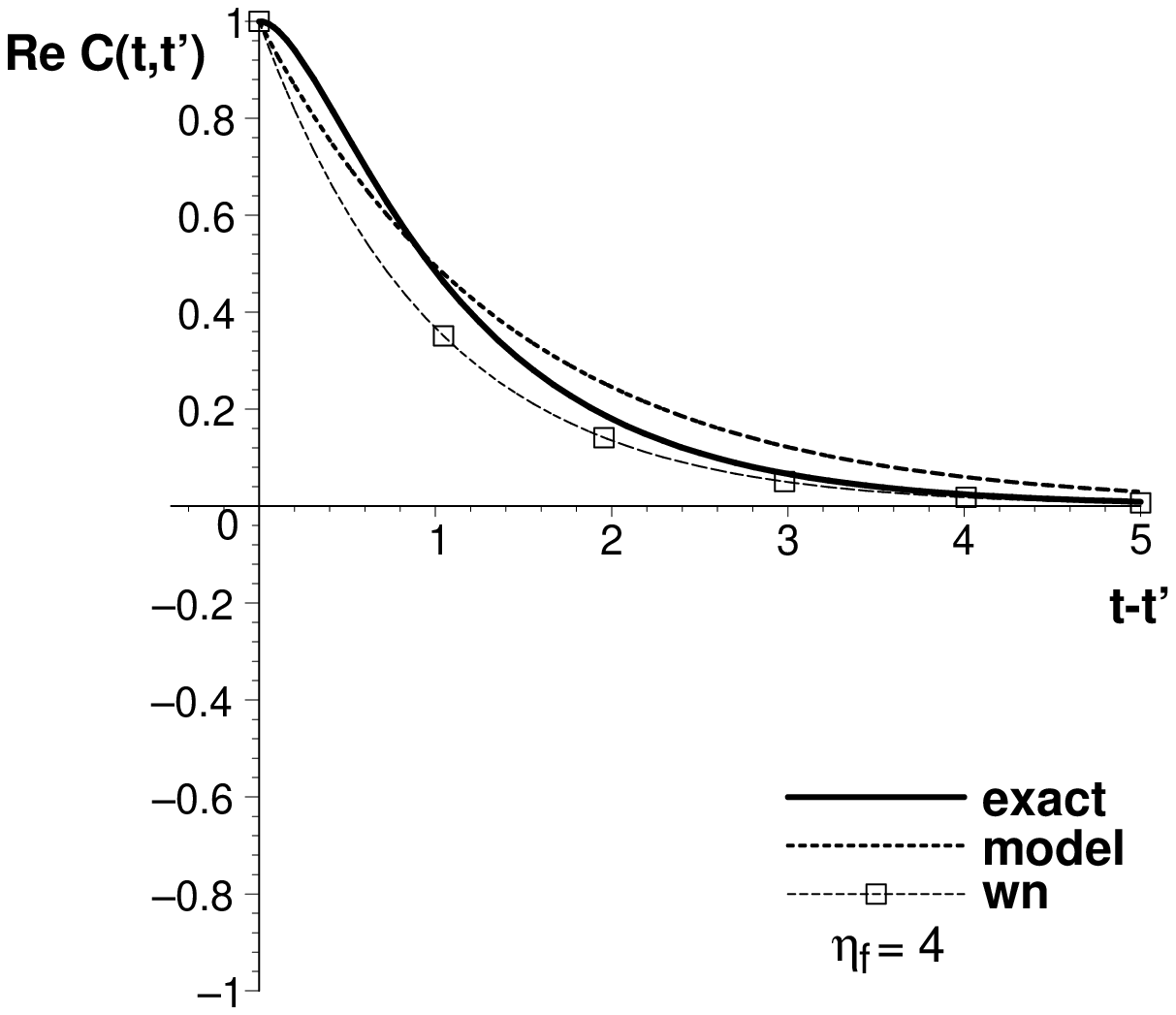, height=2.0in, 
     bbllx=100,bblly=240,bburx=465,bbury=555} \\
\epsfig{file=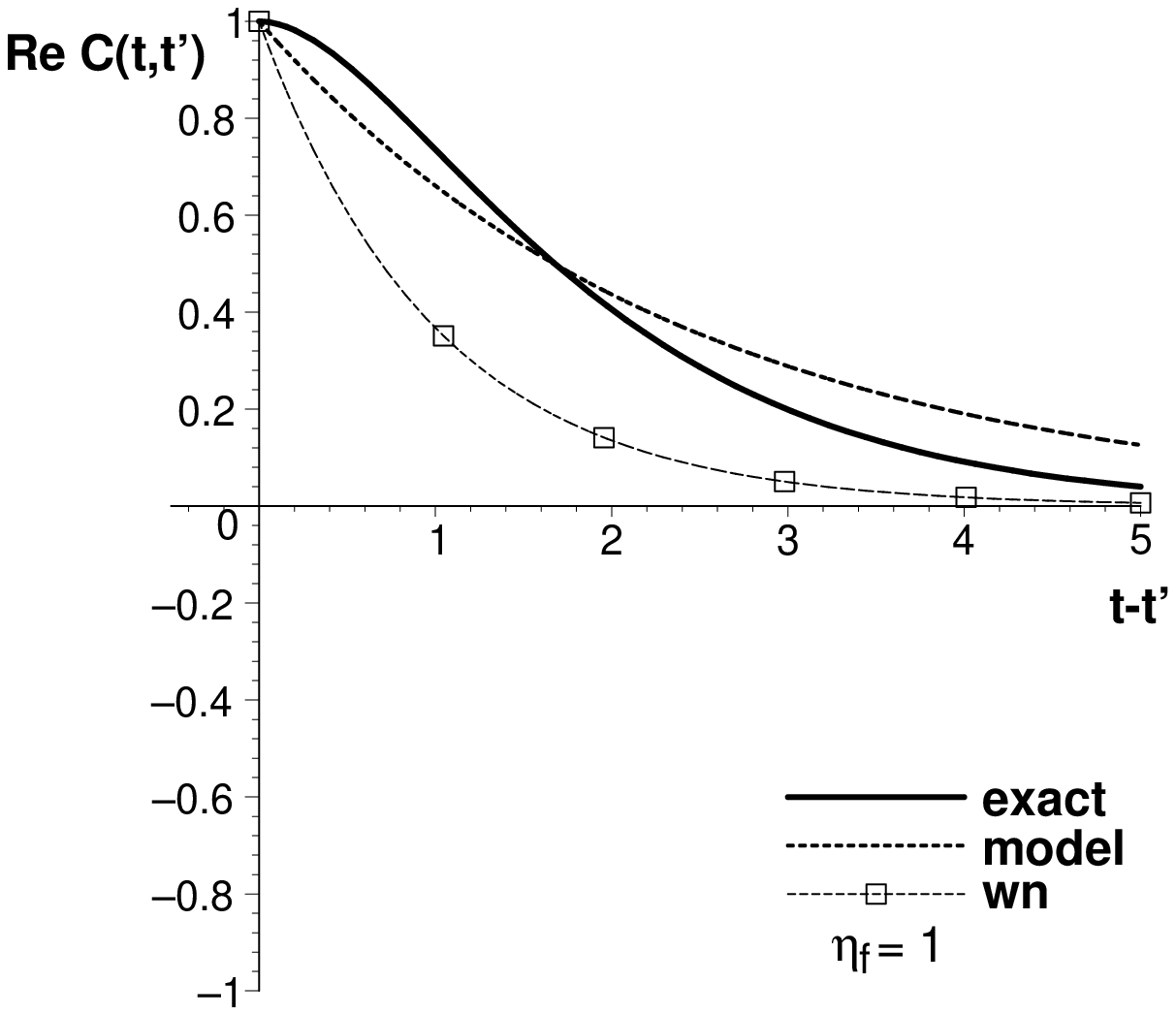, height=2.0in, 
     bbllx=100,bblly=240,bburx=465,bbury=555} \\
\epsfig{file=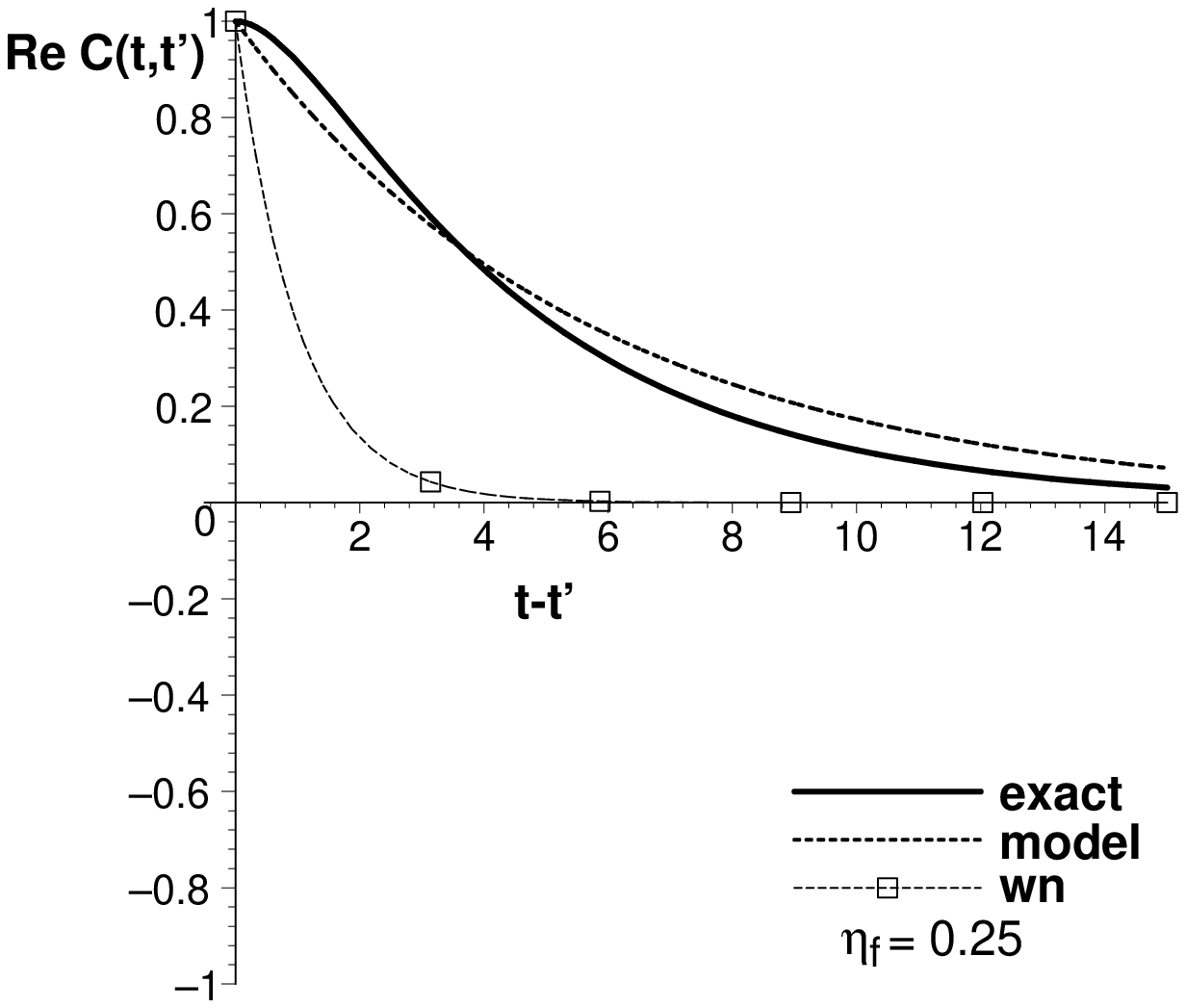, height=2.0in, 
     bbllx=065,bblly=240,bburx=465,bbury=555}
\end{center}
\caption{$\Re C(t,t')/C_0$ \It{vs.} $t-t'$, for the exact Langevin result
of \Eq(Langevin-ss-C2), for the Multiple-Rate model with decorrelation
rate $\eta_C$ given by \Eq(eta_c-ss), and for the simple white-noise
assumption $C(t,t')=C_0 \exp(-\eta |t-t'|)$.  Time is normalized such that
$\eta=1$, and the value of $\eta_f$ is noted in each figure.}
\end{figure}



\begin{figure}[p]
\begin{center}
\vspace{-0.2in}
\epsfig{file=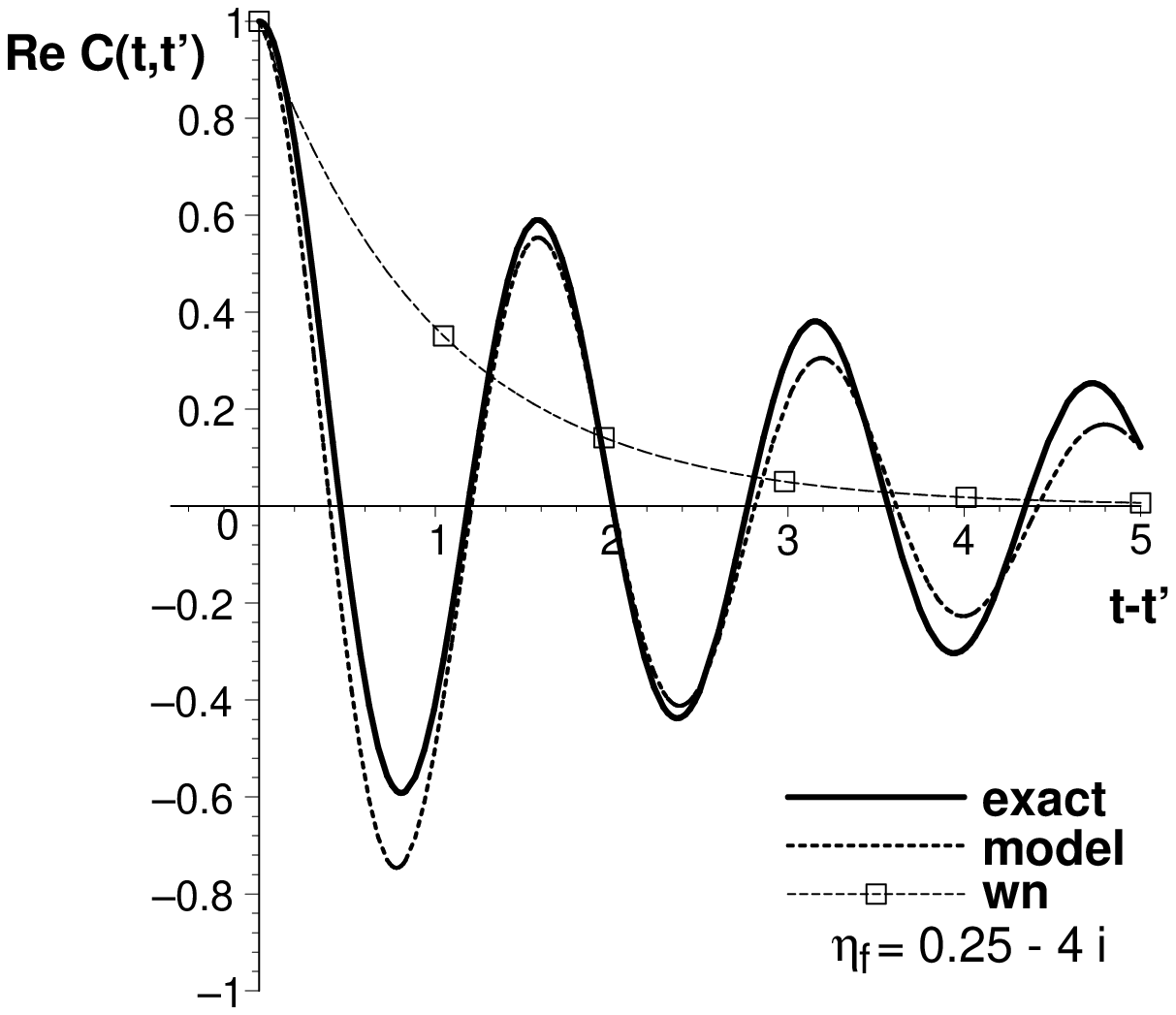, height=2.0in, 
     bbllx=100,bblly=240,bburx=465,bbury=555} \\
\epsfig{file=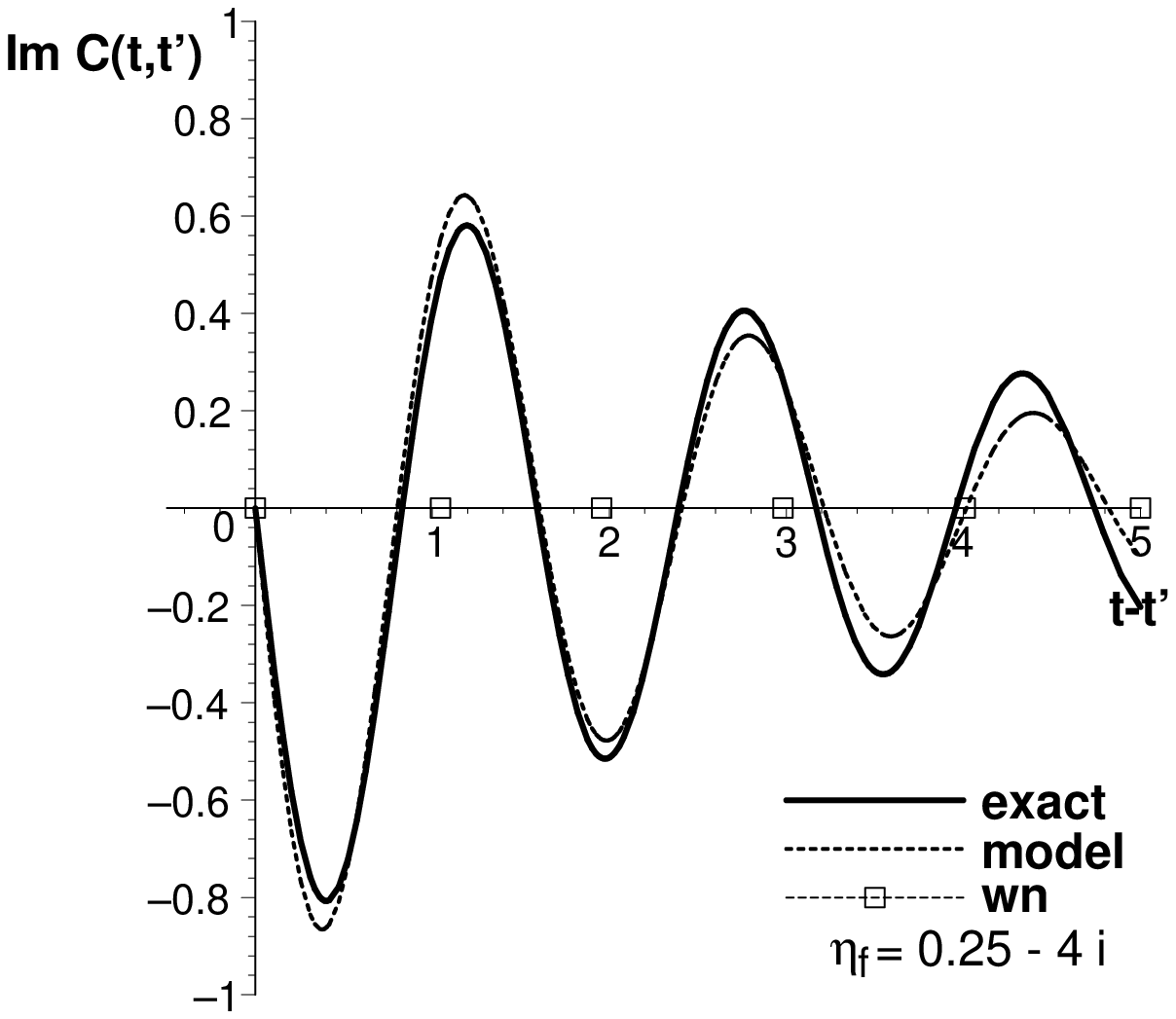, height=2.0in, 
     bbllx=100,bblly=240,bburx=465,bbury=555} \\
\end{center}
\vspace{-0.1in}
\caption{Real and imaginary parts of $C(t,t')/C_0$ \It{vs.} $t-t'$, for
the same three functions as in Fig. 1, but with $\eta_f=0.25 - 4
i$. Note that $\Im C=0$ for the white-noise case in this and later
figures.}
\end{figure}

\begin{figure}[p]
\begin{center}
\vspace{-0.5in}
\epsfig{file=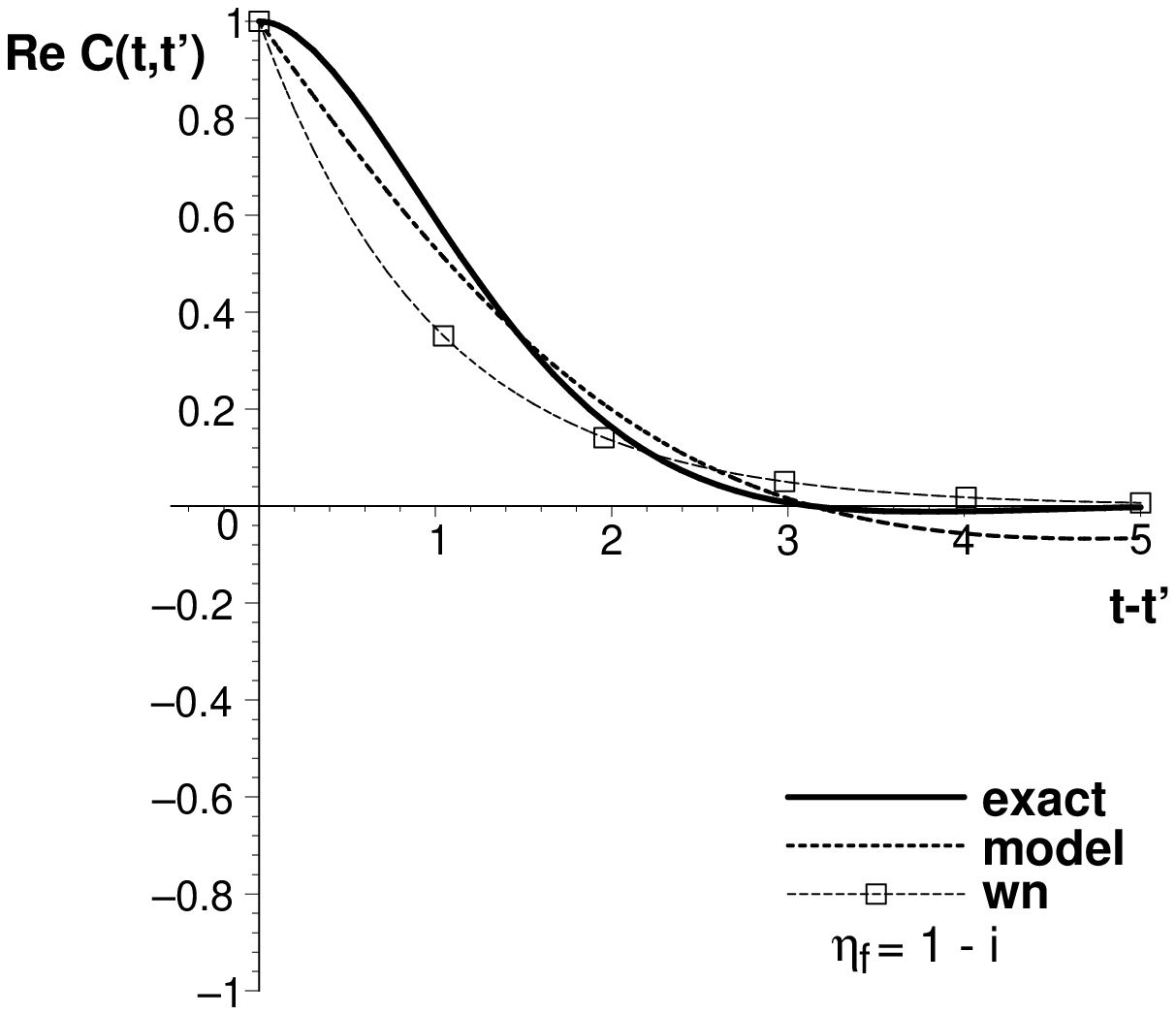, height=2.0in, 
     bbllx=100,bblly=240,bburx=465,bbury=555} \\
\epsfig{file=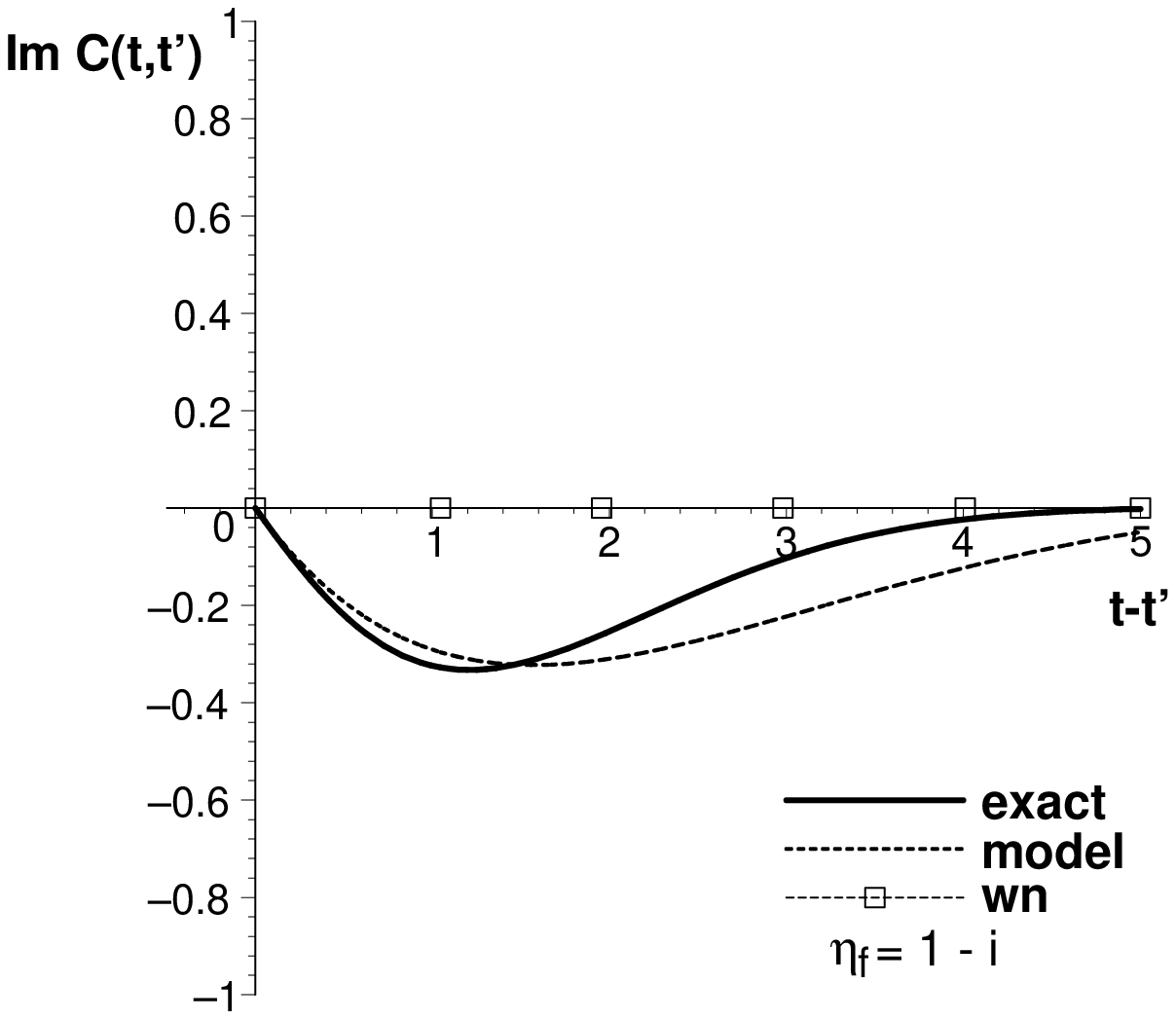, height=2.0in, 
     bbllx=100,bblly=240,bburx=465,bbury=555} \\
\end{center}
\vspace{-0.1in}
\caption{Real and imaginary parts of $C(t,t')/C_0$ \It{vs.} $t-t'$, for
the same three functions as in Fig. 1, but with $\eta_f=1 - i$.}
\end{figure}


\begin{figure}[p]
\begin{center}
\vspace{-0.2in}
\epsfig{file=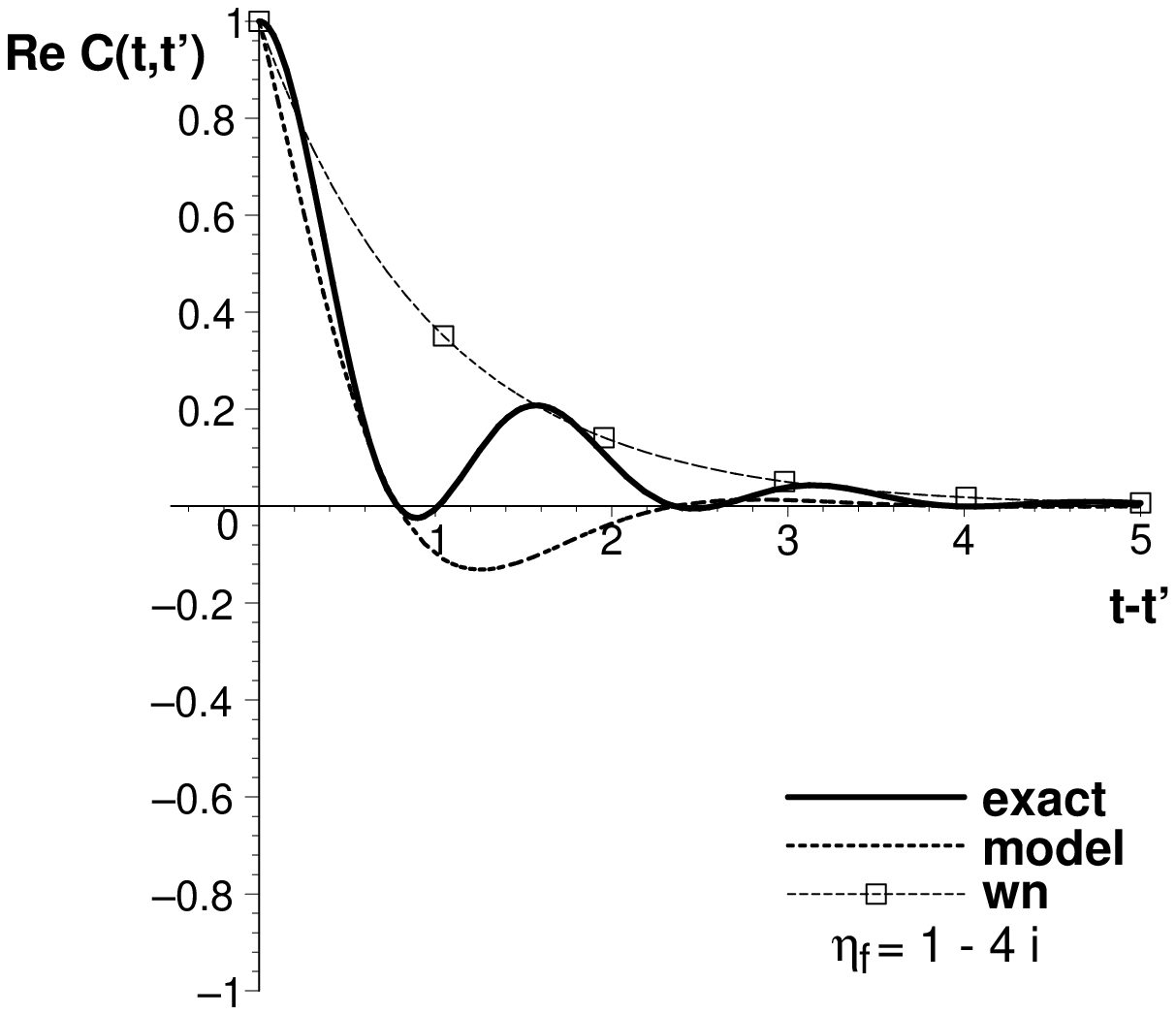, height=2.0in, 
     bbllx=100,bblly=240,bburx=465,bbury=555} \\
\epsfig{file=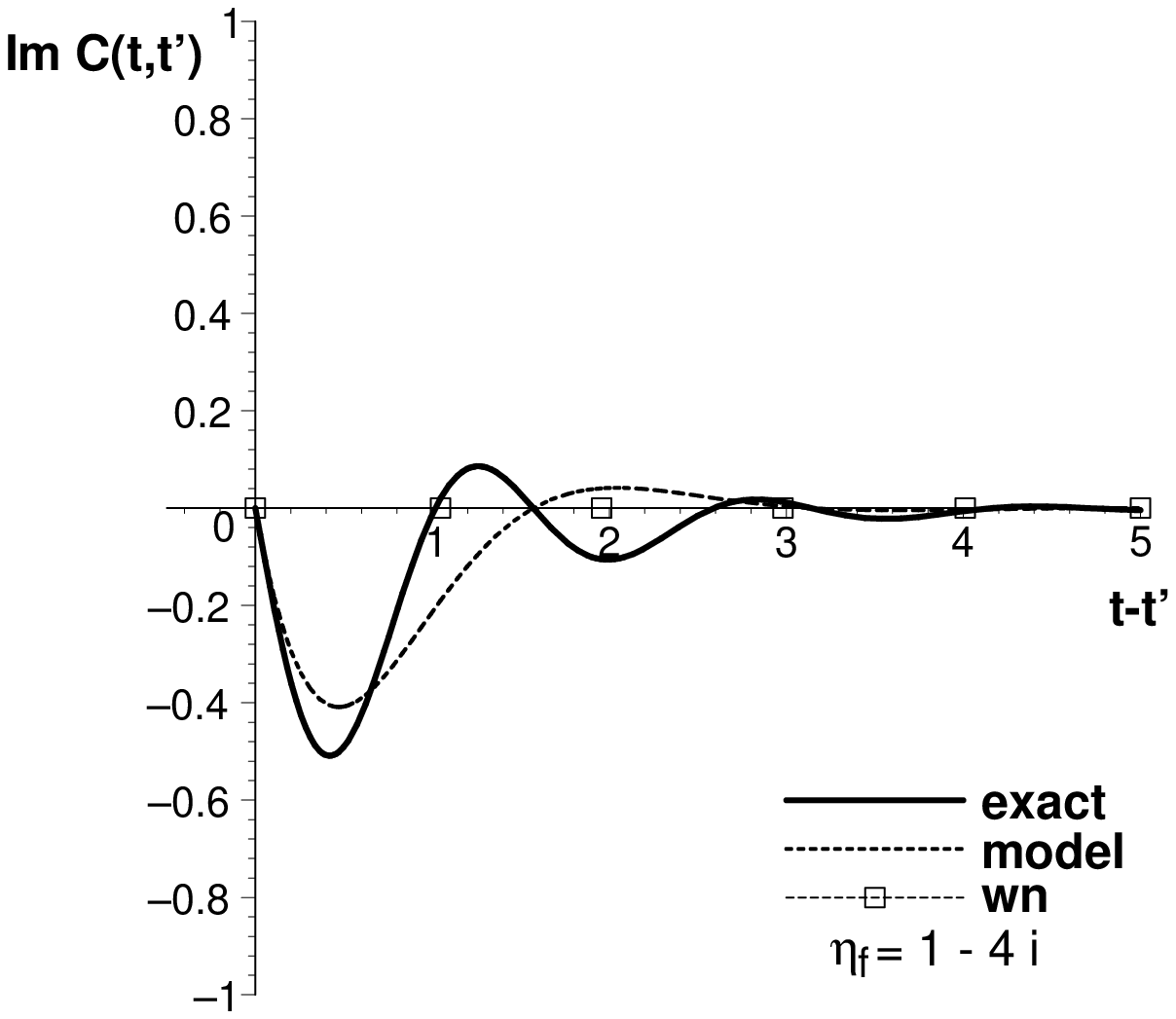, height=2.0in, 
     bbllx=100,bblly=240,bburx=465,bbury=555} \\
\end{center}
\vspace{-0.1in}
\caption{Real and imaginary parts of $C(t,t')/C_0$ \It{vs.} $t-t'$, for
the same three functions as in Fig. 1, but with $\eta_f=1 - 4 i$.}
\end{figure}

\begin{figure}[p]
\begin{center}
\vspace{-0.5in}
\epsfig{file=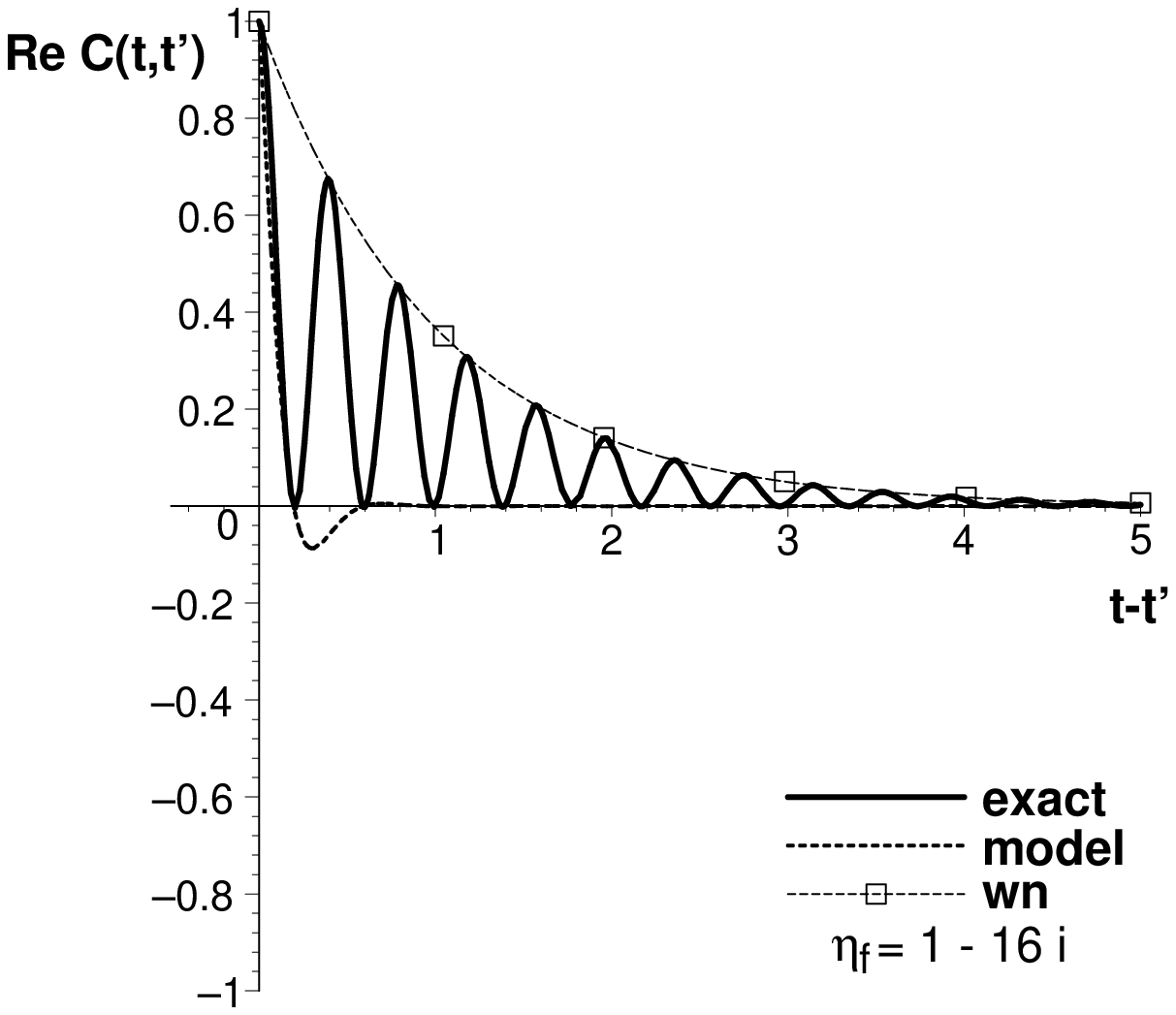, height=2.0in, 
     bbllx=100,bblly=240,bburx=465,bbury=555} \\
\epsfig{file=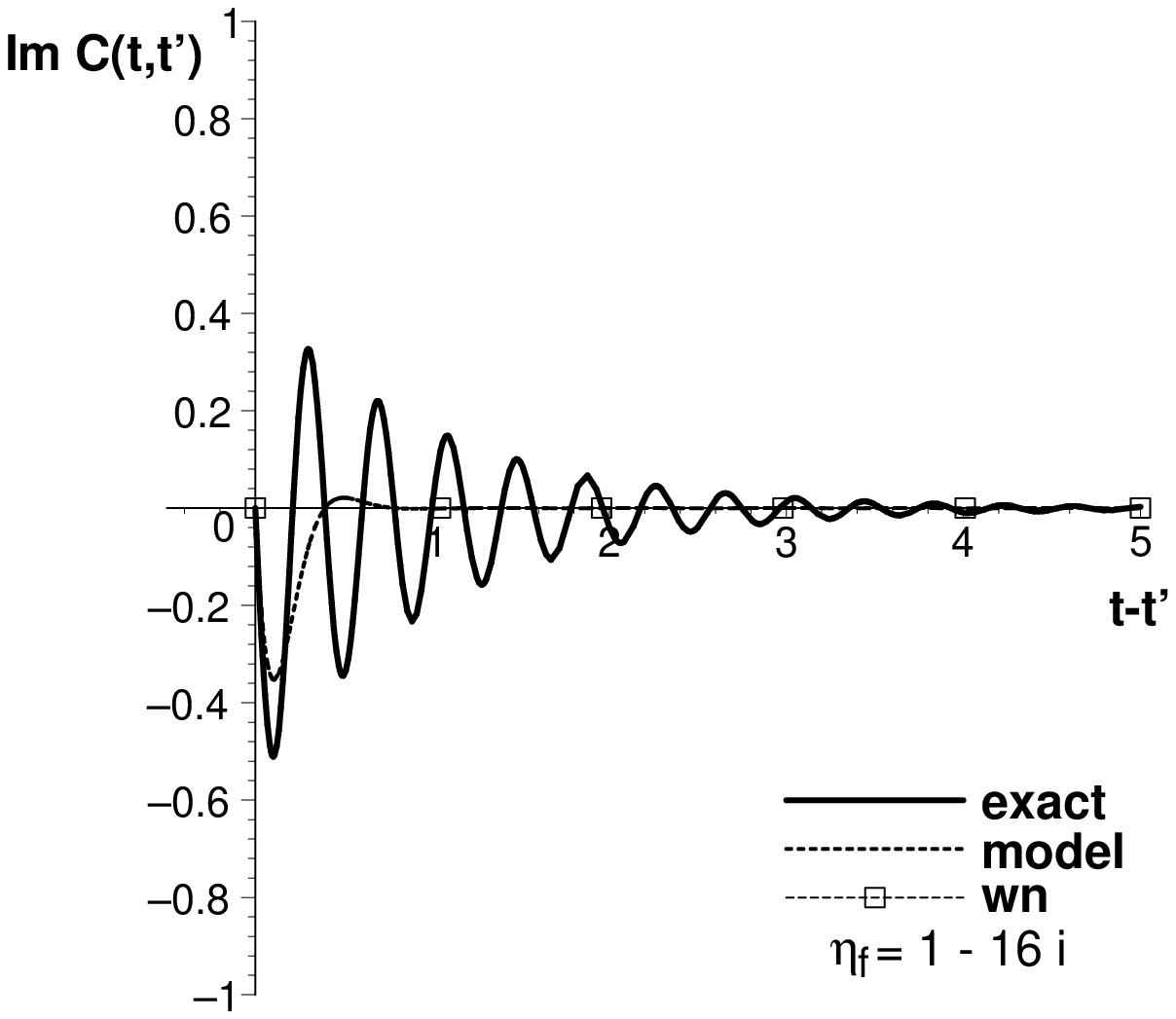, height=2.0in, 
     bbllx=100,bblly=240,bburx=465,bbury=555} \\
\end{center}
\vspace{-0.1in}
\caption{Real and imaginary parts of $C(t,t')/C_0$ \It{vs.} $t-t'$, for
the same three functions as in Fig. 1, but with $\eta_f=1 - 16 i$.}
\end{figure}

\begin{figure}[!ht]
\begin{center}
\vspace{-0.2in}
\epsfig{file=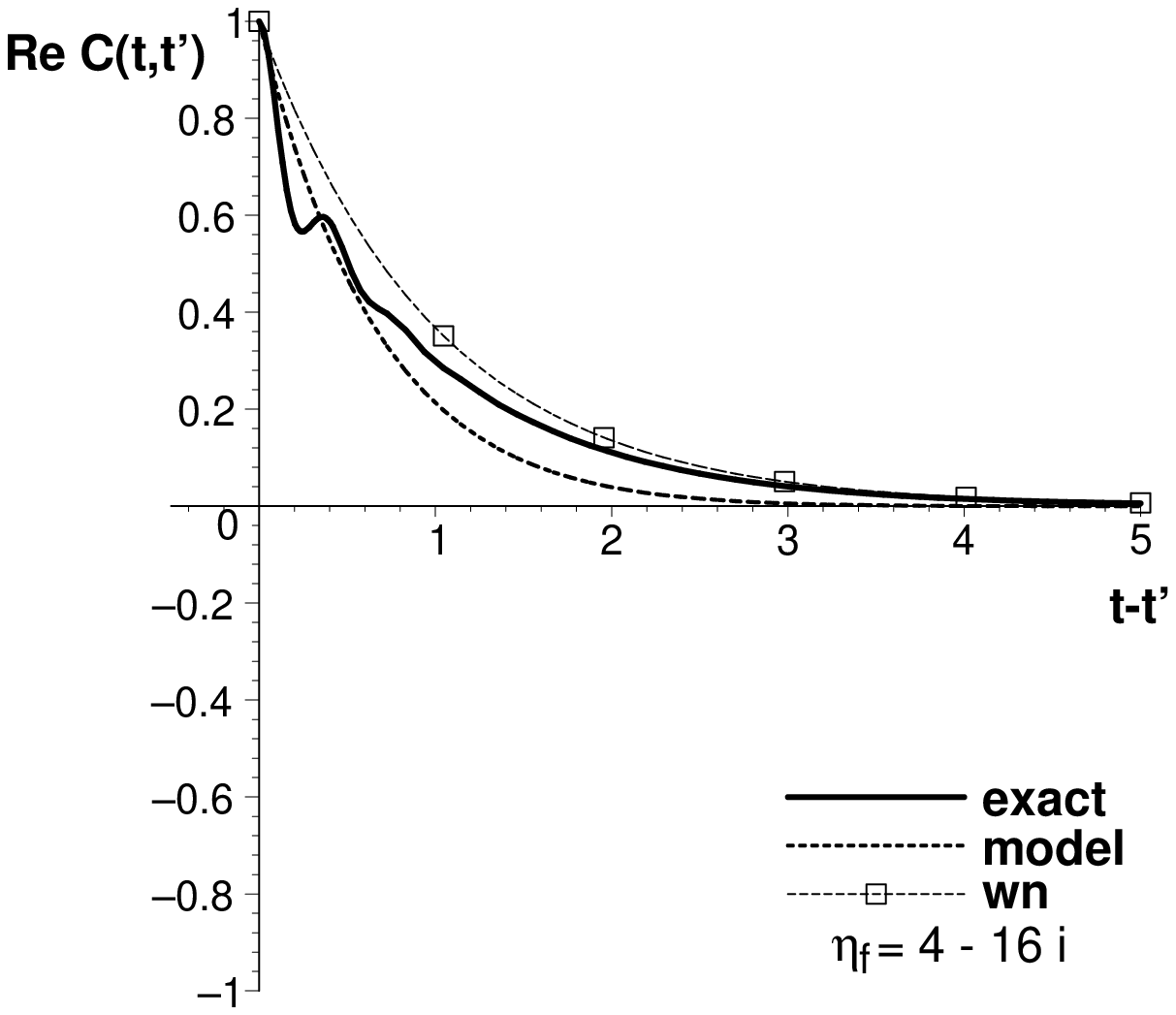, height=2.0in, 
     bbllx=100,bblly=240,bburx=465,bbury=555} \\
\epsfig{file=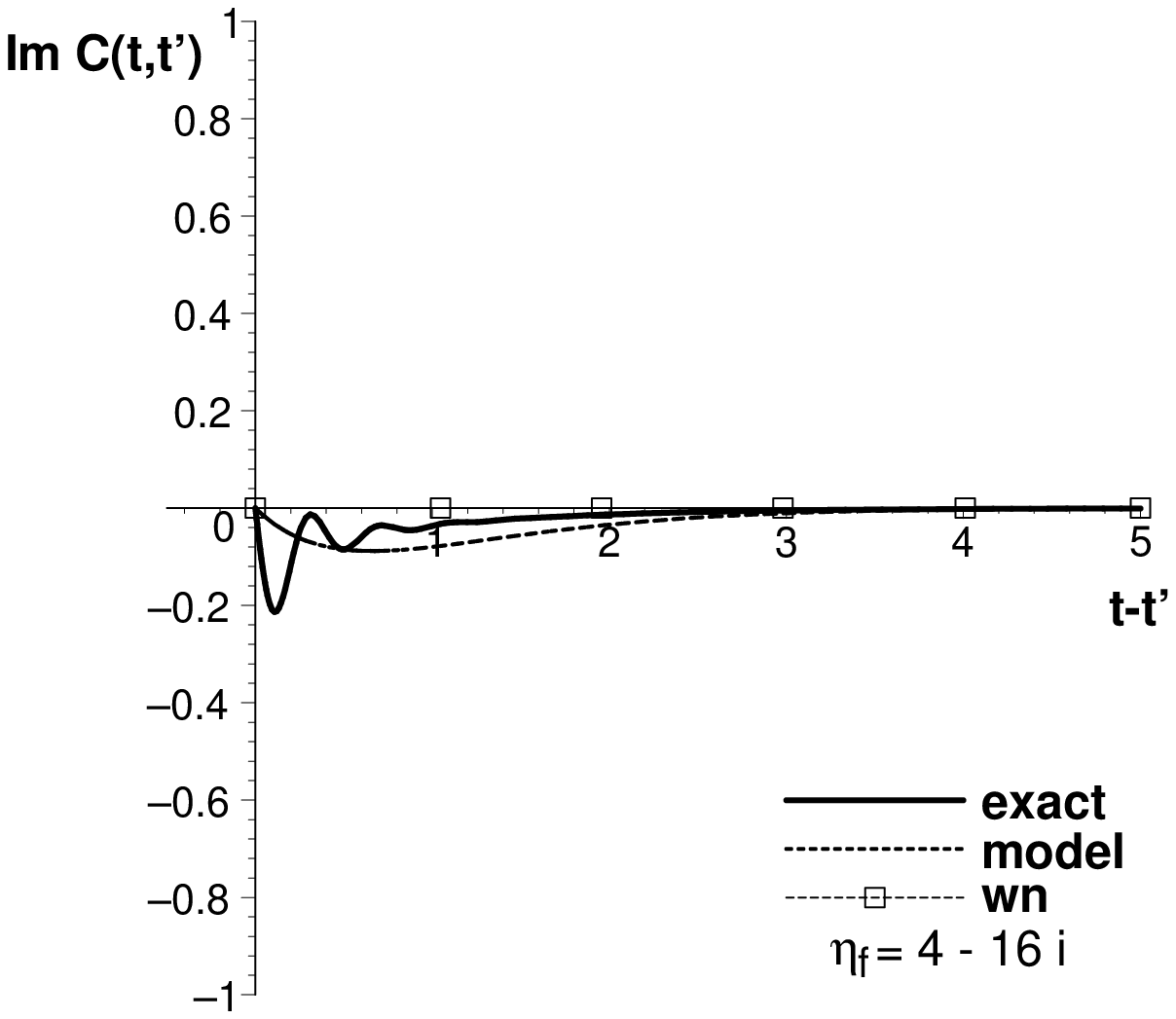, height=2.0in, 
     bbllx=100,bblly=240,bburx=465,bbury=555} \\
\end{center}
\vspace{-0.1in}
\caption{Real and imaginary parts of $C(t,t')/C_0$ \It{vs.} $t-t'$, for
the same three functions as in Fig. 1, but with $\eta_f=4 - 16 i$.}
\end{figure}

\subsection{Comparison of the Multiple-Rate model with exact Langevin result}

Figs.~(1-6) provide a comparison of the exact and model results for various
parameters.  The exact Langevin solution for $C(t,t')/C_0$ is given by
\Eq(Langevin-ss-C2).  The curves labeled ``model'' are for the
Multiple-Rate Markovian
model $C_{\mod}(t,t')/C_0 = \exp(-\eta_C |t-t'|)$, where $\eta_C$ is
obtained by solving \Eq(eta_c-ss).  The curves labeled ``wn'' are the results
for a simple white-noise assumption $C(t,t')/C_0=\exp(-\eta |t-t'|)$.
The plots show both the real and imaginary parts of $C(t,t')$, except when
$C(t,'t)$ is purely real.

The results are shown in Figs.~(1-6) for a variety of parameters.  We choose
$\eta = 1$ as a standard normalization in all cases.  Only the frequency
mismatch ($\Delta \omega = \Im (\eta - \eta_f^*)$) between the oscillator
and the random driving term and the relative
decorrelation rate ($\Re(\eta_f) / \Re(\eta)$) can matter.  Thus we
choose a frame of reference
where $\Im \eta = 0$ and any frequency mismatch is reflected in the value
of the noise frequency $\Im \eta_f$.

These comparisons show that the non-white-noise Multiple-Rate model for 
$\eta_C$ does fairly well in most cases.
All formulas of course agree well in the white-noise limit of $\Re \eta_f
\gg \Re \eta$.
The errors of the white-noise model are particularly large in the
``red-noise limit'' $\Re \eta_f \ll \Re \eta$, though they are noticeable
even if $\Re \eta_f \sim \Re \eta$.  The white-noise model has a purely real
correlation function in all cases, thus missing the frequency
shifts that arise when $\Im \eta_f \ne 0$, while the multiple-rate model
does a fairly good job
of capturing the real and imaginary parts of $C(t,t')$ in most cases.  The
most challenging case for even the multiple-rate 
model is depicted in Fig.~(5),
where there is a large frequency mismatch but comparable decorrelation
rates, $\Re \eta_f \sim \Re \eta$.  However, \Eq(Langevin-C0) shows that
the amplitude, $C_0 \sim 2 C_{f0} /(\Delta \omega)^2 \sim 2 C_{f0} /
(\Im(\eta-\eta_f^*))^2$, will be small in this strongly non-resonant case,
and perhaps does not matter much compared to resonant interactions in
realistic many-mode turbulence cases.  
Strongly non-resonant cases are easier to model with disparate values of
$\Re \eta_f$ and $\Re \eta$, as shown in Fig.~(6) and Fig.~(2), because
interference effects are less important.
To do better for the non-resonant case with $\Re \eta_f \sim \Re \eta$
would probably require
a more elaborate two-exponential model than \Eq(Cmod-ss), to allow for the
constructive and destructive interference effects represented in Fig.~(5).
Of course, for the simple Langevin case of this section, such a
model could exactly reproduce \Eq(Langevin-ss-C2), although for more
complicated cases it would again become a model to be fit to the true
$C(t,t')$ dynamics.
(Another approach, which might improve the long-time fit a bit, might be
to use $C_{\mod}^*(t,t')(t-t')$ as the weight function in
\eqr{etac-invariant} instead of just $C_{\mod}^*(t,t')$.)



\subsection{Time-dependent Langevin statistics}
\label{Sec-Time-dependent-Langevin}

We now return our attention to the more general Langevin problem with
time-dependent $\eta(t)$ and time-varying statistics for the
noise term $f^*(t)$.  That is, for generality, we also allow the noise
amplitude (given by the equal-time covariance $C_{f}(t) \doteq C_f(t,t)$) and
the noise decorrelation rate to vary in time.  Our choice of a
self-consistent model for $C_f(t,t')$ to accomplish this is motivated by
BKO's demonstration that the following form is a realizable correlation
function:
\begin{equation}
C_f(t,\bar{t}) = C_f^{1/2}(t) \exp\left[-\int_{\bar{t}}^t dt'' \,
\eta_f(t'')
\right] C_f^{1/2}(\bar{t})
\label{Langevin-noise-mod}
\end{equation}
(for $t \ge \bar{t}$).  [BKO show this is realizable as long as
$\Re(\eta_f(t)) \ge 0$ almost everywhere.] Using this expression,
\Eq(Langevin-1time) can be written as 
\begin{equation}
{\partial C(t) \over \partial t} + 2 \Re \eta(t) C(t) 
= 2 \Re C_f^{1/2} (t) \Theta^* (t),
\label{Langevin-NWM-C}
\end{equation}
where
\begin{equation}
\Theta(t) \doteq \int_0^t d \bar{t}\, R(t,\bar{t}) 
 \exp \left[-\int_{\bar{t}}^t \, dt'' \eta_f(t'') \right]
C_f^{1/2}(\bar{t}).
\label{Langevin-theta-int}
\end{equation}
Taking the time derivative of this expression, and using 
\Eq(Langevin-R), leads to
\begin{equation}
{\partial \Theta(t) \over \partial t} = - [\eta(t) + \eta_f(t)] \Theta(t) +
C_f^{1/2}(t),
\label{Langevin-NWM-theta}
\end{equation}
which is more convenient to use in a time-dependent calculation than
\Eq(Langevin-theta-int).  The initial condition is $\Theta(0)=0$.
\Eq(Langevin-NWM-C) and \Eq(Langevin-NWM-theta) can be used to determine the
equal-time covariance $C(t)$, but how can we determine the decorrelation
rate $\eta_C$ from the two-time correlation function $C(t,t')$?  [In the
full nonlinear equations used for the DIA, $\psi$ for one mode appears in
noise terms for other modes, and so we would like to know the decorrelation
rate as well as the amplitude $C(t)$.]  Even in the steady-state limit of
the previous section, we found that the full two-time correlation function
$C(t,t')$ had a more complicated form than a simple exponential, and so we
fit a simpler model $C_{\mod}(t,t')$ to it in order to determine an
effective decorrelation rate $\eta_C$.

We follow a similar procedure here.  We again use BKO's form for a
realizable time-dependent two-time correlation function to provide a model
of $C(t,t')$,
\begin{equation}
C_{\mod}(t,t') = C^{1/2}(t) \exp\left[-\int_{t'}^t dt'' \,
\eta_C(t'')
\right] C^{1/2}(t')
\label{Langevin-Cmod}
\end{equation}
(for $t \ge t'$).  Consider the integral
\begin{equation}
A(t) = \int_0^t dt' \, C_{\mod}^*(t,t') C(t,t').
\label{Langevin-Area}
\end{equation}
This is the time-dependent analog of \Eq(etac-invariant).  Rather than try
to use this to determine $\eta_C$ directly, it is more convenient to
again take time derivatives.  If $C(t,t')$ in \Eq(Langevin-Area) is
replaced with $C_{\mod}(t,t')$ of \Eq(Langevin-Cmod), then
\begin{equation}
{\partial A(t) \over \partial t} = C^2(t) -[\eta_C(t) + \eta_C^*(t)] A +
{1 \over C(t) } {\partial C(t) \over \partial t} A .
\label{Langevin-Adot1}
\end{equation}
If we instead calculate $\partial A / \partial t$ with the full $C(t,t')$
in \Eq(Langevin-Area), and use \Eq(Langevin-2time) to evaluate $\partial 
C(t,t') / \partial t$, then
\begin{eqnarray}
{\partial A(t) \over \partial t} & = & C^2(t) -[\eta(t) + \eta_C^*(t)] A 
  + {1 \over 2 C(t) } {\partial C(t) \over \partial t} A   \nonumber \\
  & & + \, \Theta_3^*(t) C^{1/2}(t) C_f^{1/2}(t),
\label{Langevin-Adot2}
\end{eqnarray}
where
\begin{equation}
\Theta_3^*(t) = \int_0^t d t' \, {C_{\mod}^*(t,t') \over C^{1/2}(t)} 
   \int_0^{t'} d \bar{t} \,
R^*(t',\bar{t}) { C_f^*(t,\bar{t}) \over C_f^{1/2}(t)} .
\label{Langevin-Bint}
\end{equation}
Taking the time derivative of this, and using \Eq(Langevin-noise-mod) for
$C_f(t,\bar{t})$, gives
\begin{equation}
{\partial \Theta_3(t) \over \partial t} = C^{1/2}(t) \Theta(t) - [\eta_C(t)
+ \eta_f(t)] \Theta_3(t).
\label{Langevin-Theta3dot}
\end{equation}

Equating \Eq(Langevin-Adot1) and \Eq(Langevin-Adot2), one can then solve for
the effective decorrelation rate $\eta_C$.  Using \Eq(Langevin-NWM-C)
to eliminate the $\partial C(t)/\partial t$ term, the result is
\begin{eqnarray}
\eta_C(t) & = & {\cal P} \left( \eta(t) - \Re \eta(t) + {C_f^{1/2}(t) \Re
\Theta(t) \over C(t) } \right. \nonumber \\
  & & \left. - \, {\Theta_3^*(t) C^{1/2}(t) C_f^{1/2}(t) \over A(t)} \right),
\label{Langevin-eta_c}
\end{eqnarray}
where we have added the ${\cal P}$ operator to enforce realizability for
the reasons discussed below. Here ${\cal P}(z) = z$ if $\Re z \ge 0$ and ${\cal
P}(z)=i \Im z$ if $\Re z < 0$.  Substituting \Eq(Langevin-NWM-C) into
\Eq(Langevin-Adot1) gives
\begin{eqnarray}
{\partial A(t) \over \partial t} & = & C^2(t) -2 \Re (\eta(t) +\eta_C(t))
   A(t) \nonumber \\
   & &  + \, 2 \Re \Theta(t) {C_f^{1/2}(t) \over C(t)} A(t).
\label{Langevin-Adot}
\end{eqnarray}
%
%
%
%
Eqs.~(\ref{Langevin-NWM-C}), (\ref{Langevin-NWM-theta}), and
(\ref{Langevin-Theta3dot}-\ref{Langevin-Adot}) provide a complete set of
equations that can be integrated forward in time.  They comprise a
Markovian closure theory (including non-white noise effects) for the 
time-dependent Langevin equation.  The relevant initial conditions are
discussed below.  This set of equations can be used to determine the
amplitude $C(t)$ and the effective decorrelation rate
$\eta_C(t)$ used to model the two-time behavior $C(t,t')$.

In a normal long-time statistical steady state, where $\eta$, $\eta_f$
and $C_f$ are constants (and $\Re(\eta)>0$ and $\Re(\eta_f)>0$), then one
can show that the second and third terms on the right-hand side of
\Eq(Langevin-eta_c) cancel and that it reproduces the steady-state result
for $\eta_C$ in \Eq(Langevin-eta_c-ss).

Consider the behavior of these equations in an unstable case, with $\Re
\eta = -\gamma < 0$.  For simplicity, assume the coefficients $\eta$,
$\eta_f$ and $C_f$ are all constant in time, with $\Re(\eta_f)>0$.  Then
one can show that $C(t)$ eventually grows as $\exp(2 \gamma t)$,
while $\Theta(t) \sim \exp((\gamma - \eta_f) t)$ grows more slowly, so that
the third term on the right-hand side of \Eq(Langevin-eta_c) vanishes.  The
fourth term on the right-hand side of \Eq(Langevin-eta_c) also vanishes
because $\Theta_3 \sim \exp(2 \gamma t)$ while $A \sim \exp( 4 \gamma t)$.
In this limit, $\eta_C = \eta - \Re(\eta)$.

Thus with constant coefficients, the two cases of positive or negative $\Re
\eta$ will, at least in the long-time limit, naturally reproduce the
limiting operator ${\cal P}(\eta) = \Re \eta H(\Re \eta) + i \Im \eta$,
which was introduced by BKO\cite{Bowman93} to preserve realizability for
the assumed form of $C(t,t')$ in \Eq(Langevin-Cmod).  In the white-noise
limit $\eta_f \gg \eta$, it is straightforward to show that
realizability is ensured for all time, not just
in the long-time limit (see also Appendix~(\ref{Appendix-real})).
These results might suggest that the $\cal P$ operator in
\Eq(Langevin-eta_c) is not needed, if its argument 
always has a positive real part anyway.  However, by numerically
integrating Eqs.~(\ref{Langevin-NWM-C}), (\ref{Langevin-NWM-theta}), and
(\ref{Langevin-Theta3dot}-\ref{Langevin-Adot}), we have found cases
where this is not true and the ${\cal P}$ operator is needed in
\Eq(Langevin-eta_c) to enforce the realizability condition $\Re \eta_C\ge 0$.
[Without the $\cal P$ operator, $\Re \eta_C$ will transiently go negative
in some strongly non-resonant cases such as $\eta=1$ and
$\eta_f=0.25+16i$.]  Eqs.~(\ref{Langevin-NWM-C}, \ref{Langevin-NWM-theta})
are an exact system of equations for the equal time covariance $C(t)$ for
Langevin dynamics, which ensures that $C(t)$ is always positive.  But
according to Theorem 2 of BKO\cite{Bowman93} (and
Appendix~\ref{Appendix-real} of the present paper), $\Re \eta_C \ge 0$ is
necessary for $C_{\mod}(t,t')$ as given by \Eq(Langevin-Cmod) to be a
realizable two-time correlation function.  This may be important if $C_{\rm
mod}(t,t')$ is in turn used in a noise term driving some other Fourier
mode.

Formally, the initial conditions for this system of equations require some
care to handle an apparent singularity, but in practice this should not be a
problem.  With a finite initial $\psi(0)$ in \Eq(Langevin-solution), the
initial conditions for the Markovian closure equations are $\Theta(0)=0$,
$A(0)=0$, $\Theta_3(0)=0$, and $C(0)=C_1$.  For short times, we then have
$C(t) \approx C_1$, $\Theta(t) \approx C_f^{1/2} t$.  If $\eta_C$ is
finite, then for short times we also have $A(t) = C_1^2 t$ and $\Theta_3 =
(C_1 C_f)^{1/2} t^2/2$.  It follows from \Eq(Langevin-eta_c) that
$\eta_C=\eta-\Re \eta + t C_f/(2 C_1)$ for short times, which is a consistent
solution that is finite and continuous, resolving the $0/0$ ambiguity in
the last term of \Eq(Langevin-eta_c).  In a numerical code, it is
convenient to use the initial conditions $\Theta(0)=0$, $\Theta_3(0)=0$
(thus assuming the initial noise $C_f=0$), $C(0)=C_1$, and $A(0)=C_1^2
\Delta t$, where $\Delta t$ is a time step smaller than any other
relevant time scales in the problem.

%


\section{Formulation of the full nonlinear problem and statistical
closures}\label{closures}

In this section we provide background on the general form of the nonlinear
problem we are considering and on the general theory of statistical
closures.  In particular we will write down Kraichnan's
direct-interaction approximation, which is the starting point of our
calculation.  This section borrows heavily from the BKO
paper\cite{Bowman93} (including some of their wording), but is provided for
completeness to define our starting point.  

\subsection{The fundamental nonlinear stochastic process}

Consider a quadratically nonlinear equation, written in Fourier space,
for some variable~$\psik$:
\begin{equation}
\linop\psik(t) = \half\SD \Mkpq\psip^*(t)\psiq^*(t). \eq(master)
\end{equation}
Here the \It{time-independent} coefficients of linear ``damping''~$\nuk$
and mode-coupling~$\Mkpq$ may be complex. Given random initial conditions,
we seek ensemble-averaged (or, if the system is ergodic, time-averaged)
moments of $\psik(t)$, taking for simplicity the mean value of $\psik$ to
be zero.

Many important nonlinear problems can be represented in this form with a
simple quadratic nonlinearity.  For example, the two-dimensional
Navier--Stokes equation for neutral fluid turbulence can be written
in this form, where $\psi$ represents the stream function such
that the velocity ${\v v} = \hat{{\v z}} \times \grad \psi$, and $\Mkpq =
\hat{{\bf z}} \cdot \vp \times \vq (q^2 - p^2) / k^2$.
Other examples include Charney's barotropic vorticity equation for planetary
fluid flow, and a class of two-dimensional plasma drift wave turbulence
problems (such as the Hasegawa--Mima equation or the Terry--Horton
equation).  Some
three-dimensional one-field plasma turbulence problems can also be written in
this form since the dominant $\vec E \times \vec B$ nonlinearity acts only
in two dimensions perpendicular to the magnetic field.  The three-dimensional
Navier--Stokes equations and general multi-field plasma turbulence equations
can also be written in the form of \Eq(master) if $\psik$ is considered as
a vector and $\nuk$ and $\Mkpq$ become matrices or tensors.  In fact,
BKO\cite{Bowman93} consider covariant multiple-field formulations of the
DIA and Markovian closures.  Here we will focus on the one-field case, where
$\psik$ is a scalar amplitude for mode $\vk$.

For each $\vk$ in \Eq(master), the summation on the right-hand-side
involves a sum over all possible $\vp$ and $\vq$ that satisfy the
three-wave interaction $\vk + \vp +\vq = 0$ (this is sometimes expressed as
$\vk = \vp + \vq$, but the reality conditions $\psi_{-\vk} = \psik^*$
has been used to rearrange it).  Without any loss of generality one may
assume the symmetry 
\begin{equation}
\Mkpq=\Mkqp.  \eq(Msym)
\end{equation}
Another important symmetry possessed by many such systems
is
\begin{equation}
\sigk\Mkpq+\sigp\Mpqk+\sigq\Mqkp=0	\eq(Menergysym)
\end{equation}
for some time-independent nonrandom \It{real} quantity~$\sigk$.  [See
Refs.~\onlinecite{Armstrong62} and~\onlinecite{Sagdeev69} for the relation
between this symmetry and the Manley-Rowe relations for wave actions.]
\Equation(Menergysym) is easily shown to imply that the nonlinear
terms of \Eq(master) conserve the ensemble-averaged total
\It{generalized energy} $E \doteq\half\sum_\vk\sigk \langle
\Abs{\psik(t)}^2 \rangle$. [The nonlinear terms also conserve the
generalized energy in each individual realization, although we will be
focusing on ensemble-averaged quantities, where $\langle \ldots
\rangle$ denotes ensemble-averaging.]
For some problems, \Eq(Menergysym) may be satisfied by more than one choice
of~$\sigk$; this implies the existence of more than one nonlinear
invariant.  For example, in the case of two-dimensional hydrodynamics,
\Eq(Menergysym) is satisfied for both $\sigma_\vk = k^2$
and $\sigma_\vk = k^4$, which correspond to the conservation of energy and
enstrophy, respectively.


We define the \It{two-time correlation function}~$\Ck(t,t')
\doteq \<\psik(t)\psik^*(t')>$ and the \It{equal-time correlation function}
$\Ck(t)\doteq\Ck(t,t)$ (note that the two functions are
distinguished only by the number of arguments), so that~$E=\half\sum_\vk
\sigk\Ck(t)$.  In stationary turbulence, the two-time correlation function
depends on only the difference of its time arguments:
$\Ck(t,t')\doteq {\cal C}_\vk(t-t')$.
The renormalized \It{infinitesimal response function} (nonlinear Green's
function)~$\Rk(t,t')$ is the ensemble-averaged infinitesimal response to a
source function~$S_\vk(t)$ added to the right-hand side of \Eq(master) for
mode $\vk$ alone.
As a functional derivative, 
\begin{equation}
\Rk(t,t') \doteq \left. \< {\delta \psik(t) \over \delta S_\vk(t')}> 
\right|_{S_\vk=0}.
\end{equation}
%
%
We adopt the convention
that the equal-time response function $\Rk(t,t)$ evaluates to $1/2$
[although $\lim_{\epsilon \goesto 0+}$ $ \Rk(t+\epsilon,t)=1$].


\subsection{Statistical closures; the direct-interaction approximation}

The starting point of our derivation will be the equations of Kraichnan's
direct-interaction approximation (DIA), as given in Eqs.~(6-7) of
BKO,\cite{Bowman93} and reproduced below as
Eqs.~(\ref{closure}-\ref{DIA}).

%
The general form of a statistical closure in the absence of mean
fields is 
\begin{mathletters}\eq(closure)
\begin{eqnarray}
\linop\Ck(t,t') &+& \I0t d\tb\,\sk(t,\tb)\Ck(\tb,t')\nonumber\\
&=&\I0{t'}d\tb\,\cFk(t,\tb)\Rk^*(t',\tb), \eq(closure a)
\end{eqnarray}
\begin{eqnarray}
\linop\Rk(t,t') &+& \I{t'}t d\tb\,\sk(t,\tb)\Rk(\tb,t')\nonumber\\
&=& \Dirac{t-t'}.
\eq(closure b)
\end{eqnarray}
\end{mathletters}%
While these equations (with the expressions for $\sk$ and $\cFk$ given
below) are an approximate statistical solution to \Eq(master),
they are the exact statistical solution to a generalized
Langevin equation
\begin{equation}
\left( {\partial \over \partial t} + \nu_\vk \right) \psi_\vk(t) 
+ \int_0^t d \bar{t}\, \Sigma_\vk(t,\bar{t}) \psi_\vk(\bar{t}) = f_\vk(t),
\label{generalized-Langevin}
\end{equation}
where $\Sigma_\vk$ is the kernel of a non-local damping/propa\-ga\-tion
operator, and $\cFk(t,\bar{t}) = \langle f_k(t) f_k^*(\bar{t}) \rangle$.  
These equations specify an initial-value problem for which~$t=0$
is the initial time.

The original nonlinearity in \Eq(master) gives rise to two types of terms
in \Eqs(closure): those describing nonlinear damping ($\sk$) and one
modeling nonlinear noise ($\cFk$). 
%
The nonlinear damping and noise in
\Eqs(closure) are determined on the basis of fully nonlinear statistics.

The direct-interaction approximation provides specific \It{approximate}
forms for~$\sk$ and~$\cFk$:
\begin{mathletters}\eq(DIA)
\begin{equation}
\sk(t,\tb)=-\SDtext \Mkpq\Mpqk^*\Rp^*(t,\tb)\Cq^*(t,\tb), \eq(DIA a)
\end{equation}
\begin{equation}
\cFk(t,\tb)=\half\SD \Abs{\Mkpq}^2\Cp^*(t,\tb)\Cq^*(t,\tb).\eq(DIA b)
\end{equation}
\end{mathletters}%
These renormalized forms can be obtained from the formal perturbation
series by retaining only selected terms.
While there are infinitely many ways of obtaining a renormalized expression,
Kraichnan\cite{Kraichnan61} has shown that most of the resulting closed
systems of equations lead to physically unacceptable solutions.  For
example, they might predict the physically impossible situation of a
negative value for~$\Ck(t,t)$ (\ie., a negative energy)!  Such behavior
cannot occur in the DIA or other realizable closures.

The DIA also conserves all of the same generalized energies
($\half\sum_\vk\sigk \Abs{\psik(t)}^2$) 
that are conserved by the primitive dynamics.
To show this important property, it is useful to write
the equal-time covariance equation in the form
\begin{mathletters}\eq(DIACeqBTh)
\begin{equation}
\delt\Ck(t) + 2\Re\Nk(t) = 2 \Re \Fk(t), \eq(DIACeqBTh a)
\end{equation}
where
\begin{equation}
\Nk(t)\doteq \nuk\Ck(t)-\SD\Mkpq\Mpqk^*\bar{\Theta}_\pqk^*(t),
\end{equation}
\begin{equation}
\Fk(t)\doteq \half \SD\Abs{\Mkpq}^2\bar{\Theta}_\kpq^*(t),
\end{equation}
\begin{equation}
\bar{\Theta}_\kpq(t)\doteq \I{t_0}t d\tb\,\Rk(t,\tb)\,\Cp(t,\tb)\,\Cq(t,\tb),
\eq(DIACeqBTh d)
\end{equation}
\end{mathletters}%
given initial conditions at the time $t=t_0$
(unless otherwise stated, we will take $t_0=0$).
As shown in BKO,\cite{Bowman93} the symmetries~\hide\Eqs(Msym)
and~\hide\Eq(Menergysym) ensure that \Eq(DIACeqBTh a) conserves all
quadratic nonlinear invariants of the form
$E\doteq\half\sum_\vk\sigk\Ck(t)$ in the dissipationless case
where $\Re\nuk=0$.\label{conservation}
The Markovian closures that BKO\cite{Bowman93} developed, and that we
extend here, preserve the structure of Eqs.~(\ref{DIACeqBTh}) and so have
all of the same quadratic nonlinear conservation properties as the original
equations. [One can show that $\Fk$ is always real, so the $\Re$ operation
on $\Fk$ in \Eq(DIACeqBTh a) is redundant.]

The DIA equations~\hide\Eq(closure) and~\hide\Eq(DIA) provide a closed set
of equations, but are fairly complicated because they involve convolutions
over two-time functions.  Their general numerical solution requires ${\cal
O}(N_t^3)$ operations, or $ {\cal O}(N_t^2)$ operations in steady state.
As described in BKO\cite{Bowman93} and Krommes,\cite{Krommes2002} a
Markovian approximation seeks to simplify this complexity by parameterizing
the two-time functions in terms of a single decorrelation rate.  Our approach
here is essentially to generalize this to allow several rate parameters to
be used, to allow the decorrelation rate for $C_\vk(t,t')$ to differ from
the decay rate for $R_\vk(t,t')$.


\section{Response functions in a statistical steady state}
\label{sec-Markov-ss}

Markovian models provide approximations that can simplify the integrals
in \Eqs(closure).
For insight, we will first investigate the long-time limit where a
statistical steady-state should be reached, so that the two-time
correlation function $C(t,t')$ and response function $R(t,t')$ can depend
only on the time difference $t-t'$.  
%
%
%
In a statistical steady state, all of the Markovian models
in BKO\cite{Bowman93} use a simple exponential behavior for
$\Ck(t,t')$ and $\Rk(t,t')$.  Here we will assume the model forms
\begin{equation}
R_{\mod,\vk}(t,t') = \exP{-\nk(t-t')} H(t-t')     \eq(Rkexp-ss)
\end{equation}
and
\begin{equation}
C_{\mod,\vk}(t,t')\doteq\cases{ C_{0\vk} \exP{-\nck(t-t')}     
                              &for~$t\ge t'$,\cr 
C_{0\vk} \exP{-\nck^*(t-t')} &for~$t < t'$.\cr}
                                      \eq(Ckexp-ss)
\end{equation}
Note that $\nk$ is the decay rate for the infinitesimal response function
$\Rk$, while $\nck$ is the decorrelation rate for $\Ck(t,t')$.

Inserting \Eq(DIA a) into \Eq(closure b) and using the exponential forms
of \Eq(Rkexp-ss) and \Eq(Ckexp-ss) in the integrals yields
\begin{eqnarray}
&& \linop\Rk(t,t') = \Dirac{t-t'} \nonumber \\
&+& \SD \Mkpq \Mpqk^*C_{0\vq} H(t-t') \nonumber \\
&& \times \int_{t'}^t d \bar{t} \,
 \exP{-(\np^* + \ncq^*) (t-\bar{t})
      -\nk(\bar{t}-t') }.  \eq(closure-ss b2)
\end{eqnarray}
Evaluating the integral gives
\begin{eqnarray}
&& \linop\Rk(t,t') = \Dirac{t-t'} \nonumber \\
&+& \SD {\Mkpq \Mpqk^*C_{0\vq} \over \np^* + \ncq^* -\nk} H(t,t') \nonumber \\
&\times& \left[ \exP{-\nk(t-t')} -\exP{-(\np^*+\ncq^*)(t-t')} 
       \right]. \eq(closure-ss b3)
\end{eqnarray}
The solution to this equation for $t>t'$ is
\begin{eqnarray}
&& \Rk(t,t') = \exP{-\nuk (t-t')} \nonumber \\
&& + \SD {\Mkpq \Mpqk^* C_{0\vq} \over \np^* + \ncq^* -\nk} \nonumber \\
&& \times \left[ {\exP{-\nuk (t-t')} - \exP{-\nk (t-t')} \over
\nk -\nuk } \nonumber \right. \\
&& \left. - {\exP{-\nuk (t-t')} - \exP{-(\np^* + \ncq^*) (t-t')} \over
\np^* + \ncq^* -\nuk} \right].  \eq(Rk-ss)
\end{eqnarray}
Clearly this is not strictly consistent with the simple exponential form
for $\Rk$ assumed in \Eq(Rkexp-ss) and used to evaluate the integrals in
\Eq(closure b).  We will instead fit the model \Eq(Rkexp-ss) to
\Eq(Rk-ss), in the same way that we did in the Langevin case for
\Eq(etac-invariant).  Requiring that both \Eq(Rkexp-ss) and the full
\Eq(Rk-ss) give the same weighted average over time (where $R_{{\rm
mod},\vk}^*$
is used as the weight to ensure invariance to frequency shifts) gives
\begin{eqnarray}
&& {1 \over \nk + \nk^*} \doteq \int_{t'}^{\infinity} dt \,
R_{\mod,\vk}^*(t,t') \Rk(t,t').
\eq(Rk-ss-int)
\end{eqnarray}
Inserting \Eq(Rk-ss) on the right-hand side, and carrying out a few lines
of algebra, the result is
\begin{eqnarray}
&& {1 \over \nk +\nk^*} = {1 \over \nuk + \nk^*}  \nonumber \\
&&\quad + \SD {\Mkpq \Mpqk^* C_{0\vq} \over (\nuk +\nk^*) (\nk +\nk^*) 
       (\np^* + \ncq^* +\nk^*)}.
\end{eqnarray}
A little rearranging leads to
\begin{equation}
\nk \doteq \nuk - \SD {\Mkpq \Mpqk^* C_{0\vq} \over \nk^* + \np^* + \ncq^* }.
\eq(etak-ss)
\end{equation}
Note that this has a similar form to the steady-state decay rate in the
DIA-based EDQNM, such as in Eq.~(39b) of BKO\cite{Bowman93} (but with
their $\nq^*$ replaced by $\ncq^*$).


One can go through a similar calculation of $\Ck(t,t')$, and calculate its
weighted time average to determine the decorrelation rate $\nck$.
We will not do so now, as one can instead just take the steady-state limit
of the results in the next section.  \Eq(etak-ss) can also be obtained from
the steady-state limit of the results in the next section, and so provides
a useful cross-check.  

We note that there is some flexibility in the choice of weighting in
\Eq(Rk-ss-int).  We could use $C_{\mod, \vk}^*(t,t')$ as the weight
instead of $R_{\mod, \vk}^*(t,t')$.  Either choice preserves Galilean
invariance.  Using this alternate weight, \Eq(Rk-ss-int) becomes
\begin{eqnarray}
&& {C_{\vk 0} \over \nk + \nck^*} \doteq \int_{t'}^{\infinity} dt \,
C_{\mod,\vk}^*(t,t') \Rk(t,t')
\eq(Rk-ss-intC)
\end{eqnarray}
and the resulting expression for $\nk$ is like \Eq(etak-ss) but with
$\nk^*$ on the right-hand side of \Eq(etak-ss) replaced by $\nck^*$,
which would automatically agree with the steady-state
$\bar{\eta}_k$ to be defined in \Eq(MRMC-ss).  But it turns out that the
main steady-state results of Sec.~(\ref{Markov-properties}) hold with
either choice of weights, and it seems more symmetric and makes
more sense as a standard fitting procedure to use $R_{\mod,\vk}^*$ as the
weight for integrating $R_\vk$ in \Eq(Rk-ss-int).
%
%
This raises the question of whether to use $C_{\mod,\vk}^*$ or
$R_{\mod,\vk}^*$ as the weight function for time averages of
$C_\vk(t,t')$, as we will do in the next section.
%
%
We can resolve this ambiguity by going back to the steady-state
Langevin problem of Sec.~(\ref{Sec-Langevin-ss}).  If one tries to use
$R^*(t,t')$ as the weight in
\Eq(etac-invariant), so that it becomes
\begin{equation}
{C_0 \over \eta_{C} + \eta^*} \doteq \int_{-\infinity}^t dt' \,
\exp(-\eta^*(t-t')) C(t,t'),
\label{etac-invariant2}
\end{equation}
then one can go through the same steps used to derive \Eq(eta_c-real) and
find that in the limit of real coefficients it gives $\eta_C = \eta
\eta_f / (2 \eta + \eta_f)$.  In the red-noise limit $\eta_f \ll \eta$, this
gives $\eta_C = \eta_f/2$, which is a factor of 2 off from the correct
result ($\eta_C = \eta_f$) for the red noise limit.  
Thus, we will use
$C_{\mod,\vk}^*(t,t')$ as the weight for taking time-averages of
$C_\vk(t,t')$ and use $R_{\mod,\vk}^*(t,t')$ for time-averaging
$R_\vk(t,t')$.
The weighting choices might be 
reconsidered in a multi-field generalization of a Markovian closure,
where the requirement of 
covariance may impose constraints on the choice of the weight functions,
but it seems that the symmetric choices made here are most likely to
generalize well.
%

\section{Time-Dependent Multiple-Rate Markovian Closure}
\label{sec-Markov}

Applying these techniques in a straightforward way to the time-dependent
DIA equations leads to the Multiple-Rate Markovian Closure (MRMC)
equations. The two-time correlation function is modeled with the
realizable form
\be
C_{\mod, \vk}(t,t') = C_\vk^{1/2}(t) C_\vk^{1/2}(t') \exp\left(-\int_{t'}^t d
\bar{t} \, \nck(\bar{t})\right)
\label{Cmodk}
\ee
(for $t>t'$, with $C_{\mod, \vk}(t,t')=C_{\mod, \vk}^*(t',t)$ for
$t<t'$), and the response function is modeled as
\be
R_{\mod, \vk}(t,t') = \exp\left(-\int_{t'}^t d \bar{t} \, \nk(\bar{t})\right) H(t-t').
\label{Rmodk}
\ee
Denoting $\bar{\Theta}_{\vk \vp \vq}(t) = \Theta_{\vk \vp \vq}(t)
C_p^{1/2}(t) C_q^{1/2}(t)$, and inserting Eqs.~(\ref{Cmodk}-\ref{Rmodk})
into \Eq(DIACeqBTh d), we can write the equal-time DIA covariance equations
of \Eq(DIACeqBTh) as
\begin{mathletters}\eq(MRMC)
\begin{equation}
\delt\Ck(t) + 2\Re\bar{\eta}_\vk(t)\,\Ck(t) = 2 \Fk(t), \eq(MRMC a)
\end{equation}
\begin{equation}
\bar{\eta}_\vk \doteq 
           \nuk-\SD \Mkpq\Mpqk^* \Thpqk^*(t)\, \Cq^{1/2}(t) \Ck^{-1/2}(t),
           \eq(MRMC b)
\end{equation}
\begin{equation}
\Fk \doteq \half \SD\Abs{\Mkpq}^2\Thkpq^*(t)\,\Cp^{1/2}(t)\,\Cq^{1/2}(t),
\eq(MRMC c)
\end{equation}
\begin{equation}
\delt \Thkpq +(\nk+\ncp+\ncq)\Thkpq = C_p^{1/2}(t) C_q^{1/2}(t),
\eq(MRMC d)
\end{equation}
\begin{equation}
\Thkpq(0)=0.\eq(MRMC e)
\end{equation}
\end{mathletters}%

This is very similar to the Bowman--Krommes--Otta\-vi\-ani Realizable Markovian
Closure (RMC) (as given by Eqs.~(66a--e) of BKO\cite{Bowman93}), but with
the replacement of the single decay/decorrelation rate of the RMC with
three different rates in these equations.  [Other Markovian models, such as
the EDQNM closure, also use a single decorrelation rate parameter.] If in
\Eq(MRMC d) we replace $\eta_\vk=\bar{\eta}_\vk$, $\ncp={\cal
P}(\bar{\eta}_\vp)$, and $\ncq={\cal P}(\bar{\eta}_\vq)$, then these
equations become identical to the RMC.

To summarize the three different rates used here: 
\begin{itemize}

\item $\bar{\eta}_\vk$ is the nonlinear energy damping rate for the wave
energy equation for the equal-time covariance $C_k(t)$ in \Eq(MRMC a), and
is defined in \Eq(MRMC b);

\item $\nk$ is the decay rate for the infinitesimal response function
$R_k(t,t')$ in \Eq(Rmodk), and is defined in \Eq(eta_k);

\item and $\nck$ is the decorrelation rate for $C_k(t,t')$ in \Eq(Cmodk),
and is defined in \Eq(eta_Ck).
\end{itemize}

To determine $\nk(t)$ and $\nck(t)$, we follow a similar procedure as we
did for the time-dependent Langevin equation in
Sec.~(\ref{Sec-Time-dependent-Langevin}).  Define $A_\vk(t)$ as the
following weighted time-average of $R_\vk$
\be
A_\vk(t) = \int_0^t dt' R_{\mod, \vk}^*(t,t') R_\vk(t,t').
\eq(Ak-int)
\ee
If $R_\vk(t,t') = R_{\mod, \vk}(t,t')$ as given by \Eq(Rmodk), then
\be
{\partial A_\vk \over \partial t} = 1
- (\nk^* + \nk) A_\vk,
\eq(Akdot1)
\ee
while if $R_\vk(t,t')$ satisfies Eqs.(\ref{closure b},\ref{DIA a}), then
\begin{eqnarray}
{\partial A_\vk \over \partial t} & = &  1
- (\nk^* + \nu_\vk) A_\vk \nonumber \\
&& + \SD \Mkpq\Mpqk^* C_\vq^{1/2} \Theta_{1,\vp \vq \vk}^*,
\eq(Akdot2)
\end{eqnarray}
where
%
%
\begin{eqnarray}
&& \Theta_{1,\vp \vq \vk}^*(t) = \nonumber \\
&& \int_0^t dt' \,
{R_{\mod, \vk}^*(t,t') } 
\int_{t'}^t d \bar{t}\, 
{C_\vq^*(t,\bar{t}) \over C_\vq^{1/2}(t) }
R_\vp^*(t,\bar{t}) R_\vk(\bar{t},t').
\end{eqnarray}
It is often more convenient to work with the differential version of this,
which, after using Eqs.~(\ref{Cmodk}-\ref{Rmodk}) to replace 
$C_\vq(t,t')$ and $R_\vp(t,t')$ with their model forms, is
\be
{\partial \Theta_{1, \vp \vq \vk}^* \over \partial t} =
-(\nk^* +\ncq^* +\np^*) \Theta_{1, \vp \vq \vk}^* + C_\vq^{1/2} A_k(t)
\eq(theta1-dot)
\ee
(with the initial condition $\Theta_{1,\vk \vp \vq}(0)=0)$). Requiring that
\Eq(Akdot1) and \Eq(Akdot2) be equivalent determines $\nk$ to be
\be
\nk = \nu_\vk - {1 \over A_\vk} \SD \Mkpq \Mpqk^* C_\vq^{1/2}
\Theta_{1, \vp \vq \vk}^*.
\eq(eta_k)
\ee

The calculation of $\nck$ proceeds in a similar way.  $A_{C \vk}(t)$ is
defined as a weighted time integral of $C_\vk(t,t')$:
\be
A_{C \vk}(t) = \int_0^t dt'\, C_{\mod, \vk}^*(t,t') C_\vk(t,t').
\label{Ack-int}
\ee
If $C_\vk(t,t')$ in this integral is replaced by $C_{{\rm
mod},\vk}(t,t')$ as given by \Eq(Cmodk), then 
\be
{\partial A_{C \vk} \over \partial t} = C_\vk^2(t)
+ {1 \over C_\vk(t)} {\partial C_\vk(t) \over \partial t} A_{C \vk}
-(\nck^* + \nck) A_{C \vk},
\eq(ACkdot1)
\ee
(where we make the time dependence of $C_\vk(t)$ explicit to distinguish it
from the two-time $C_\vk(t,t')$).  If the exact dynamics for
$C_\vk(t,t')$ given by Eqs.~(\ref{closure a},\ref{DIA}) are used, then
%
%
\begin{eqnarray}
{\partial A_{C \vk} \over \partial t} & = &  C_\vk^2(t) 
+ {1 \over 2 C_\vk(t)} {\partial C_\vk(t) \over \partial t} A_{C \vk}
- (\nck^* + \nu_\vk) A_{C \vk} \nonumber \\
& + & \SD \Mkpq\Mpqk^* C_\vk^{1/2} C_\vq^{1/2} \Theta_{2, \vp \vq \vk}^*
\nonumber \\
& + & {1 \over 2} \SD |\Mkpq|^2 C_\vk^{1/2} C_\vp^{1/2} C_\vq^{1/2}
\Theta_{3,\vk \vp \vq}^*,
\eq(ACkdot2)
\end{eqnarray}
where
\begin{eqnarray}
&& \Theta_{2, \vp \vq \vk}^*(t) = \nonumber \\
&& \int_0^t dt'\, 
{C_{\mod, \vk}^*(t,t') \over C_\vk^{1/2}(t) } 
\int_0^t d \bar{t}\, 
{C_\vq^*(t,\bar{t}) \over C_\vq^{1/2}(t) }
R_\vp^*(t,\bar{t}) C_\vk(\bar{t},t')
\end{eqnarray}
and
\begin{eqnarray}
&& \Theta_{3,\vk \vp \vq}^*(t) = \nonumber \\
&& \int_0^t dt'\, 
{C_{\mod, \vk}^*(t,t') \over C_\vk^{1/2}(t) } 
\int_0^{t'} d \bar{t}\, 
{C_\vq^*(t,\bar{t}) \over C_\vq^{1/2}(t) }
{C_\vp^*(t,\bar{t}) \over C_\vp^{1/2}(t) } R_\vk^*(t',\bar{t}).
\end{eqnarray}
Using Eqs.~(\ref{Cmodk}-\ref{Rmodk}), the differential versions of these are
\bea
{\partial \Theta_{2, \vp \vq \vk}^* \over \partial t} & = &
-(\nck^* +\ncq^* +\np^*) \Theta_{2, \vp \vq \vk}^* \nonumber \\
& & +C_\vk(t) \Theta_{\vp \vq \vk}^*(t) 
  + {C_\vq^{1/2}(t) \over C_\vk^{1/2}(t)} A_{C \vk}(t) 
\eq(theta2-dot)
\eea
and
\bea
{\partial \Theta_{3,\vk \vp \vq}^* \over \partial t} & = &
-(\nck^* +\ncq^* +\ncp^*) \Theta_{3, \vk \vp \vq}^* \nonumber \\
&& + C_\vk^{1/2}(t) \Theta_{\vk \vp \vq}^*(t).
\eq(theta3-dot)
\eea
The quantity $\nck$ is then determined by equating \Eq(ACkdot1) and
\Eq(ACkdot2), yielding
\begin{eqnarray}
&& \nck = \nu_\vk 
+ {1 \over 2 C_\vk(t)} {\partial C_\vk(t) \over \partial t} \nonumber \\
&& - \, {1 \over A_{C \vk}} \SD \Mkpq\Mpqk^* C_\vk^{1/2} C_\vq^{1/2}
\Theta_{2, \vp \vq \vk}^* 
\nonumber \\
&& - \, {1 \over 2 A_{C \vk}} \SD |\Mkpq|^2 C_\vk^{1/2} C_\vp^{1/2}
C_\vq^{1/2} \Theta_{3,\vk \vp \vq}^*,
\eq(eta_Ck)
\end{eqnarray}
where \Eq(MRMC a) could be used to eliminate $\partial C_\vk(t) / \partial
t$.  As in \Eq(Langevin-eta_c) for the case of the time-dependent Langevin
equation, while there are effects in this equation that will tend to give
$\Re \nck \ge 0$, it may be necessary to modify this equation to enforce
realizability in all cases.  This is done by replacing this equation, of
the form $\nck = {\rm RHS}$, with $\nck = {\cal P}({\rm RHS})$.  Note that
it is only $\Re \nck \ge 0$ that is needed for realizability, while $\Re \nk$
can transiently go negative (as it does in two-dimensional hydrodynamics
because of the inverse cascade, or in some plasma problems where the zonal
flows may become nonlinearly
unstable\cite{Rogers2000,Krommes2000b,Diamond98}).  This is similar to
BKO's treatment.\cite{Bowman93}

The complete set of equations that constitutes the Multiple-Rate Markovian
Closure (MRMC) are Eqs.~(\ref{MRMC}) for the equal-time covariance
$C_\vk(t)$ and related quantities,
Eqs.~(\ref{Akdot1},\ref{theta1-dot},\ref{eta_k}) for quantities related to
the response function, and
Eqs.~(\ref{ACkdot1},\ref{theta2-dot}-\ref{eta_Ck}) for quantities related
to the two-time correlation function.  The MRMC extends the RMC to make
less restrictive assumptions and include additional effects, but at the
expense of a few new parameters.  In addition to replacing the single
decay/decorrelation rate of the RMC with 3 different rates, $\nk$,
$\bar{\eta}_\vk$, and $\nck$, it also replaces the single triad interaction
time of the RMC with 4 different triad interaction times,
$\Theta_{\vk \vp \vq}$,
$\Theta_{1, \vk \vp \vq}$,
$\Theta_{2, \vk \vp \vq}$, and
$\Theta_{3, \vk \vp \vq}$.
Each of these triad interaction times has a different weighting of response
functions and two-time correlation functions.  While this increases the
complexity some, the overall computational scaling of this system is
still ~${\mathcal O}(N_t)$, a significant improvement over the
${\mathcal O}(N_t^2)$ or ${\mathcal O}(N_t^3)$ scaling of the DIA.

\section{Properties of the Multiple-Rate Markovian Closure}
\label{Markov-properties}
In a steady-state limit, \Eq(eta_k) simplifies to
\begin{eqnarray}
\eta_\vk = \nu_\vk - \SD {\Mkpq\Mpqk^* C_\vq
            \over \nk^* + \eta_\vp^* + \ncq^* }.
\eq(MRMC-eta-ss)
\end{eqnarray}
%
The steady-state balance $\Re \bar{\eta}_\vk C_k = F_k$ from
\Eq(MRMC a) simplifies to
\begin{eqnarray}
&& C_\vk \Re \left[\nu_\vk - \SD {\Mkpq\Mpqk^* C_\vq
            \over \nck^* + \eta_\vp^* + \ncq^* } \right]
\nonumber \\
&& = {1 \over 2} \SD {|\Mkpq|^2 C_\vp C_\vq
            \over \eta_\vk^* + \ncp^* + \ncq^* }.
\eq(MRMC-ss)
\end{eqnarray}
[Note the subtle notational differences: the expression for $\eta_\vk$
becomes the expression for $\bar{\eta}_\vk$ if $\eta_\vk^*$ on the RHS of
\Eq(MRMC-eta-ss) is replaced by $\eta_{C \vk}^*$.]
Finally, \Eq(eta_Ck) reduces to
\begin{eqnarray}
&& \nck \doteq
           \eta_\vk - {(\nck + \nck^*)}
          \left[ \SD {\Mkpq\Mpqk^* C_\vq
            \over (\nck^*+\eta_\vp^* +\ncq^*)^2 } \right.
  \nonumber \\
&& 
         \left. + {1 \over 2 C_\vk} \SD {|\Mkpq|^2 C_\vp C_\vq \over 
            (\nck^* \! + \! \ncp^* \! +\! \ncq^*) 
         (\eta_\vk^* \! + \! \ncp^* \! +\! \ncq^*) } \right ]. \nonumber \\
\eq(MRMC eta-c)
\end{eqnarray}
Thus the decorrelation rate $\nck$ equals the response function decay
rate $\nk$ plus the two correction terms in brackets.  For the simple
steady-state non-wave case with real and positive $\eta$'s, the second
correction term will cause $\nck$ to decrease (as expected for non-white
noise), while the first term will usually have an offsetting opposite
sign and cause $\nck$ to increase.  The origin of these two terms can be
traced back to the DIA \Eq(closure a).  The second correction term
corresponds to the usual effects of non-white noise (related to the
integral involving $\cFk(t,\tb)$ in \Eq(closure a)), but the first
correction term in \Eq(MRMC eta-c) is related to the time-history
integral involving the renormalized propagator $\Sigma_\vk(t,\tb)$ in
\Eq(closure a).  Thus non-white fluctuations in other modes
$C_\vq(t,\tb)$ not only change the noise term for the $\vk$ mode, but
also change the effective damping from the time-history integral,
broadening the width of $\Sigma_\vk(t,\bar{t})$ in time (if the
fluctuations $C_\vq$ were treated as white noise, then \Eq(DIA a) would
give $\Sigma_\vk(t,\bar{t}) \propto \delta(t - \bar{t})$).  
%
%

An important property to demonstrate is that in thermal equilibrium
it is possible for these two terms to cancel exactly.  Then the
decorrelation rate and response function decay rate are equivalent, $\nck =
\nk$, and the fluctuation--dissipation theorem is satisfied.  To
demonstrate that this is true, we assume the result ($\nck = \nk$) to
simplify some of the equations and then show that this is a self-consistent
assumption.  (Note also that if $\nck = \nk$, then $\bar{\eta}_\vk =
\nk$ also.)  Splitting the first summation in brackets in \Eq(MRMC eta-c)
into two equal parts
and interchanging the $\vp$ and $\vq$ labels for one of these parts
(i.e., using an identity of the form $\Sigma G_{\vk \vp \vq} = \Sigma G_{\vk
\vp \vq}/2 + \Sigma G_{\vk \vq \vp}/2$),
the terms in brackets in \Eq(MRMC eta-c) can be written as
\begin{eqnarray}
&& {1 \over 2 C_\vk} \SD {\Mkpq \over (\nk^* + \np^* +\nq^*)^2 }
    \nonumber \\ 
&& 
\times 
(\Mpqk^* C_\vq C_\vk + M_{\vq \vp \vk}^* C_\vp C_\vk + \Mkpq^* C_p C_q )
. 
\eq(MRMC FD1)
\end{eqnarray}
In thermal equilibrium, the spectrum $C_\vk$ is given by
equipartition among modes of a generalized energy-like conserved quantity.
Consider an equipartition spectrum of the form $C_\vk = 1/ \lambda_\vk$,
where $\lambda_\vk = \sum_i \alpha^{(i)} \sigma_\vk^{(i)}$,
the $\sigma_\vk^{(i)}$ are the coefficients in \Eq(Menergysym) (related to
the quadratic invariants), and $\alpha^{(i)}$ are determined by the
initial conditions. Substituting $C_\vk = 1/\lambda_\vk$, $C_\vp =
1/\lambda_\vp$, and $C_\vq = 1/\lambda_\vq$ into \Eq(MRMC FD1), and using
Eqs.~(\ref{Msym}-\ref{Menergysym}), one can show that \Eq(MRMC FD1) indeed
vanishes, so that \Eq(MRMC eta-c) simplifies to $\nck = \nk$.  The proof
that $C_\vk = 1/\lambda_\vk$ is a solution of the steady-state \Eq(MRMC-ss)
proceeds in a similar way, interchanging the $\vp$ and $\vq$ labels for
half of the summation on the left-hand side of
\Eq(MRMC-ss), and noting that $\Re \nu_\vk=0$ in an isolated
thermal system, etc.  Rigorously, this only shows that the
equipartition spectrum $C_\vk = 1/\lambda_\vk$ is an equilibrium solution.
This paper doesn't demonstrate that it is a stable equilibrium or that
mixing dynamics will necessarily relax to this state.  For a discussion of
the Gibbs-type $H$ theorem that leads to this result, see Appendix H of
Ref.~\onlinecite{Bowman92} and Refs.\onlinecite{Kraichnan67} and
\onlinecite{Fox73}.  It is significant to note that the thermal
equilibrium result holds even if the number of modes is small, and it
does not assume that the noise spectrum is white.  This is unlike a
simple Langevin equation of the form of \Eq(Langevin) (which has a
local damping term in contrast to the time-history integral of
\Eq(generalized-Langevin)), where the two-time correlation function and
the infinitesimal response function are proportional only if the noise
is white.

We next estimate the importance of the correction terms in the
decorrelation rate for an inertial range of a turbulent steady state,
such as in two-dimensional hydrodynamics.  Typically most of the energy
is at long wavelengths ($C_\vq$ is peaked at sufficiently low $\vq$),
so that the dominant 
contributions to the sums in \Eq(MRMC eta-c) come from long wavelengths:
$|\vq| \ll |\vk|$ in the first sum, and
$|\vq| \ll |\vk|$ or $|\vp| = |\vk 
+ \vq| \ll |\vk|$ in the second sum.  This means that one can approximate
the denominators in the sums of \Eq(MRMC eta-c) using, for example,
$(\eta_{C \vk} + \eta_\vp + \eta_{C \vq}) \approx (\eta_{C \vk} +
\eta_\vk)$ (since $\eta_\vk$ and $\eta_{C \vk}$ are typically increasing
functions of $\vk$).  Similar approximations give
$(\bar{\eta}_\vk - \nu_k) \approx (\eta_\vk - \nu_k) 2 \nk^* / (\nck^* +
\nk^*)$.  Using the steady-state relation $F_\vk = \Re \bar{\eta}_\vk
C_k$ from \Eq(MRMC-ss) and the disparate scale approximations to rewrite
the second sum in \Eq(MRMC eta-c) in terms of $\bar{\eta}_\vk$, and
allowing finite dissipation but ignoring wave dynamics (so that $\nu_k$
and the various $\eta$ coefficients are real), one can show that
\Eq(MRMC eta-c) simplifies in this disparate scale limit to
\begin{eqnarray}
%
%
&& \nck \doteq
           \eta_\vk - \nu_\vk
          - {2 \nk (\nk - \nu_\vk) (\nk - \nck) \over
            (\nck + \nk)^2 }.
\label{etac-inertial}
\end{eqnarray}
This gives a cubic equation for $\nck$. For $\nu_k = 0$ the roots are
$\nck=\nk$ and $\nck = (-1 \pm \sqrt{2}) \nk$.  Our speculation is that
$\nck=\nk$ will be the usual case in a steady-state inertial range.
(This appears reasonable, but it might require numerical simulations
to test it more definitively.)  The other roots are probably unstable
equilibria, so that any perturbation away from it would eventually
approach the stable root, or may only be relevant in transient
inverse-cascade cases where $\Re\eta_\vk < 0$ ($\Re\eta_{C \vk} \ge 0 $
being required to satisfy realizability).

%
%
%
%
%
%
%
%
%
%
%
%

Thus the two correction terms in \Eq(MRMC eta-c) again exactly cancel
each other (assuming the root choice made above), leading to $\eta_{C
\vk} = \eta_\vk$ and the result that
non-white-noise corrections are asymptotically unimportant in a wide
inertial range ($\vk$ large compared to the long-wavelength
energy-containing wave number scale $\v {k_0}$).  However, this may be an
artifact of the problem that the underlying DIA, on which the MRMC is
based, does not satisfy random Galilean invariance.  As is well
known\cite{Edwards64,Kraichnan64d,Kadomtsev65,Leslie73b,Kraichnan77},
the reason the DIA predicts a slightly different spectrum ($E(k) \sim
k^{-3/2}$ in the energy cascade inertial range) than the Kolmogorov
result ($E(k) \sim k^{-5/3}$) is because of this lost random Galilean
invariance.  [The standard definitions for two-dimensional hydrodynamics
use $E(k) \sim k^3 C_{|\vk|}$ when $\psi_\vk$ represents the stream function,
so that the total energy is $\int dk \, E(k)$,
a one-dimensional integral over the magnitude of $\vk$.]  
The magnitude of this discrepancy between the DIA and dimensionally
self-similar predictions is calculated for a general equation of the form
\Eq(master) in Appendix~\ref{Appendix-scaling}.

The underlying reason for this failure of the DIA is that
the nonlinear damping and noise terms (the 
left- and right-hand sides of \Eq(MRMC-ss)) are dominated by
contributions from the energy at long wavelengths. A random-Galilean
invariant theory should depend only on the shear of longer
wavelength modes (as $\eta_\vk$ does in Orszag's phenomenological
EQDNM) and the most energetically significant interactions should occur among
comparable scales ($|\vq| \sim |\vp| \sim |\vk|$).  Then the
disparate scale approximations that led to \Eq(etac-inertial) would no
longer be valid.  In such a case,
it would seem unlikely that the two terms in \Eq(MRMC eta-c) would still
exactly cancel, and there would probably be some difference
between the decorrelation rate $\eta_{C \vk}$ and the decay rate
$\eta_\vk$.  It would therefore be interesting to try to apply the
techniques developed here (for allowing multiple rates) to other
starting equations that respect random Galilean invariance, such as the
Lagrangian-history DIA, test-field model, or renormalization group
methods.

A regime where the correction terms might not cancel each other,
and the differences between $\eta_{C \vk}$ and $\eta_\vk$ might be
significant, even with the DIA's overemphasis of long-wavelength
contributions to the eddy turnover rate, is in
ITG/drift-wave plasma turbulence, where the spectrum can 
often be anisotropic and have strong wave effects.  That is, $\nu_\vk$
can be complex, with unstable modes in some directions and damped modes
in others, so that $\nu_\vk$ and $C_\vk$ vary strongly with the direction of
$\vk$.  Some plasma cases have a reduced range of relevant nonlinearly
interacting scales, and the simplifications of disparate scales in an
inertial range used to derive \Eq(etac-inertial) are not appropriate.
The corrections might also be important in non-steady-state
transient cases (such as zonal flows with predator--prey dynamics) or
in other regimes where interactions between comparable scales dominate.
Evaluating the difference between the decorrelation rate $\eta_{C \vk}$ and
the decay rate $\eta_\vk$ in more general cases such as these
probably requires a numerical treatment.  
%
%


Finally, it is useful to demonstrate that the Multiple-Rate Markovian Closure
approximation preserves realizability, which turns out to require one
additional constraint.  The MRMC equations (\ref{MRMC}) have the underlying
Langevin equation
\begin{equation}
{\partial \psi_\vk \over \partial t} + \bar{\eta}_\vk(t) \psi_\vk(t) =
f_\vk^*(t), 
\label{underlying-Langevin}
\end{equation}
where $\bar{\eta}_\vk$ is given by \Eq(MRMC b).  The statistics that
$f_\vk$ must satisfy can be found by comparing the solution for
such a Langevin equation, given by \Eq(Langevin-1time), with
Eqs.~(\ref{MRMC}), finding the constraint $\Re \int_0^t d \bar{t}\,
\bar{R}^*(t,\bar{t}) C_f^*(t,\bar{t}) = F_k$, where $F_k$ is given by
\Eq(MRMC c) and $\bar{R}(t,\bar{t}) = \exp(-\int_{\bar{t}}^t dt''\,
\bar{\eta}_\vk(t''))$ is the propagator for \Eq(underlying-Langevin).
Using an integral form for $\Theta_\kpq$ (similar to \Eq(DIACeqBTh d)),
\[
\Theta_\kpq(t) = 
 \int_0^t d \bar{t} \, R_{\mod, \vk}(t,\tb) \, {C_{\mod,
\vp}(t,\tb) \,  C_{\mod, \vq}(t,\tb) \over C_\vp^{1/2}(t)
C_\vq^{1/2}(t) },
\]
we find that if the two-time statistics of $f_\vk$ satisfy
\begin{eqnarray*}
&& C_f(t,\bar{t}) = \exp \left[ - \int_{\bar{t}}^t d t'' 
\left( \eta_k(t'') - \bar{\eta}_\vk(t'') \right) \right] \nonumber \\
&&\quad \times {1 \over 2} \SD |\Mkpq|^2 C_{\mod,\vp}(t, \bar{t}) C_{{\rm
mod},\vq}(t,\bar{t}) 
\end{eqnarray*}
(for $t > \bar{t}$), then the MRMC is the statistical solution of
\Eq(underlying-Langevin).  As shown in Theorem 1 of Appendix B of
BKO\cite{Bowman93} (and as can be inferred from considering the
statistics of $f(t) = g(t) h(t)$, where $g$ and $h$ are statistically
independent), a product of realizable correlation functions is also a
realizable correlation function.  $C_{\mod,\vp}(t, t')$ and
$C_{\mod,\vq}(t,t')$ are individually realizable because $\Re
\eta_{C \vk}>0$ for all $\vk$.  So in order to guarantee realizability
of $C_f(t,\bar{t})$, we need to impose the additional condition that
$\Re \eta_\vk \ge \Re \bar{\eta}_\vk$.  This constraint seems physically
reasonable.  The parameter $\eta_\vk$ measures the decay rate for the
ensemble averaged response $\langle \delta \psi_\vk(t) \rangle$, which
can decay either as energy is nonlinear transferred out of mode $\vk$ or
as the energy that is in $\delta \psi_\vk$ becomes randomly phased.
The quantity $\bar{\eta}_\vk$ used in \Eq(MRMC) measures only the rate at
which net energy
(regardless of phase) is transferred out of mode $\vk$ into other modes, so
it would seem reasonable that $\bar{\eta}_\vk \le \eta_\vk$ will
naturally result.

\section{Conclusions}


In summary, we have demonstrated a method for extending Markovian
approximations of the DIA, to allow the decorrelation rate for
fluctuations to differ from the decay rate for the infinitesimal
response function (the renormalized Green's function or nonlinear
propagator). This can give a more accurate treatment of various effects
such as non-white-noise forcing terms.
In practice, the corrections to the
decorrelation rate are modest, at least in isotropic non-wave cases,
since the decorrelation rate of
the noise is usually comparable to, if not much larger than, the
decay rate for the response function.  For example, if
$\eta_f=\eta$ in the simple Langevin example of \Eq(eta_c-real), then the
decorrelation rate is $\approx 60$\% lower than its white-noise value.
Furthermore, the Multiple-Rate Markovian Closure \Eq(MRMC eta-c)
for the full DIA contains an offsetting term that can increase $\nck$,
so the net result is less clear.  This is because the DIA is related to
a generalized Langevin equation \Eq(generalized-Langevin), where
non-white fluctuations modify not only the noise term (which tends to
reduce the decorrelation rate) but also modify the renormalized
propagator in the time-history integral (which tends to increase the
decorrelation rate).

We have demonstrated that these two terms in fact exactly cancel each
other as they should in thermal equilibrium where the
fluctuation--dissipation theorem applies.  We have also found another
case, that of a wide inertial range with no waves,
where it is possible for these two corrections to offset each
other exactly, so that the 
decorrelation rate and the decay rates become equal. 
However, this may be an artifact of the loss of Galilean invariance in
the Eulerian DIA, where modes in the inertial range nonlinearly interact
predominantly with long wavelength modes.  Thus it would be interesting to try
to apply the techniques developed in this paper to other
renormalized statistical theories, in which the dominant
nonlinear interactions in an inertial range are between comparable
scales instead
of disparate scales and which properly reproduce Kolmogorov's $E(k)
\propto k^{-5/3}$ inertial-range energy spectrum instead of the
Eulerian DIA's $E(k) \propto k^{-3/2}$.
Single-rate Markovian approximations have been applied in
the past to other renormalized statistical theories\cite{McComb91}
and white-noise assumptions have also been employed in renormalization group
calculations of turbulence.\cite{McComb91} An interesting question is
whether there is some way to generalize such calculations
to allow for multiple rates and non-white noise as considered here.
Another question is whether multiple-rate extensions might modify
subgrid turbulence models.  [Such corrections would probably be important
only at short scales near the transition from resolved to unresolved
scales.]

Even in the context of an Eulerian DIA-based theory, there may be some
regimes where the multiple-rate corrections in this paper may be
important and warrant further investigation.  These might include cases
where non-steady-state dynamics are important (i.e., predator--prey
oscillations
between different parts of the spectrum, such as between drift waves and
zonal flows), or where interactions between comparable $|\vk|$ scales
are more important, such as might occur in anisotropic plasma turbulence
with wave dynamics and with instability growth rates or Landau damping
rates that vary strongly with the magnitude and direction of the
wavenumber.  One could test whether these corrections are important or
negligible in various regimes by looking at 3-mode coupling
cases,\cite{Bowman93,Bowman97} or by numerically comparing with the DIA
or direct numerical simulations.

The complete set of equations that constitutes the Multiple-Rate Markovian
Closure (MRMC) are summarized in the final paragraph of
Sec.~(\ref{sec-Markov}).  The MRMC extends the Realizable
Markovian Closure (RMC) of BKO\cite{Bowman93} to allow various nonlinear
rates and interaction times to differ.  The single
decay/decorrelation rate of the RMC is replaced with 3 different rates,
$\nk$ (the response function decay rate), $\nck$ (the decorrelation rate
for the two-time correlation function),
and $\bar{\eta}_\vk$ (the energy damping rate).  The triad interaction
time of the 
RMC is replaced with 4 different triad interaction times with various
weightings of decorrelation and decay rates.  While this
increases the complexity of the equations somewhat, the main
computational advantages of a local-in-time Markovian closure relative to the
non-local-in-time DIA are retained.

%

%

\acknowledgments
We thank Prof. John A. Krommes for many helpful discussions sharing his
insights into the DIA and Markovian approximations.  In particular, we
thank him for pointing out how to get the symmetric form of
\Eq(etac-invariant), which helps preserve important invariance properties.
This work was supported by U.S.~Department of Energy Contract
No.~DE--AC02--76CHO3073 and by the Natural Sciences and Engineering
Research Council of Canada.


\appendix

\section{Realizability of a particular two-point correlation
function}\label{Appendix-real}

In theorem 2 of their Appendix B, Bowman, Krommes, and
Ottaviani\cite{Bowman93} show one way to prove that a two-point correlation
function of the form of \Eq(Langevin-Cmod) is ``realizable'' (if $\Re
\eta_C(t) > 0$ is satisfied almost everywhere).  Realizability means
that this two-point correlation function is the exact solution to some
underlying stochastic problem, such as a Langevin equation.  In the absence
of realizability, non-physical difficulties can sometimes develop, such
as the
predicted energy $C(t) = C(t,t)$ going negative or diverging.  Here we
present an alternate proof that \Eq(Langevin-Cmod) is realizable.

Consider the standard Langevin equation with time-dependent coefficients,
but in the white-noise limit $\langle f(t) f^*(t') \rangle = 2 D(t)
\delta (t-t')$.  Then \Eq(Langevin-1time) simplifies to
\begin{equation}
{\partial C(t) \over \partial t} + 2 \Re \eta(t) C(t) 
= 2 D(t),
\label{Langevin-1time-wn}
\end{equation}
while the equation for the two-time correlation function, \Eq(Langevin-2time),
becomes just
$\partial C(t,t')/\partial t + \eta(t)C(t,t') = 0$
for $t>t'$, with the boundary condition $C(t',t')=C(t')$.  Taking the time
derivative of \Eq(Langevin-Cmod) gives
\begin{equation}
\left( {\partial \over \partial t} + \eta_C(t) \right) C_{\mod}(t,t') 
= {1 \over 2 C(t)} {\partial C(t) \over \partial t} C_{\mod}(t,t').
\end{equation}
If $C(t,t')=C_{\mod}(t,t')$, then these last two equations
give $\eta=\eta_C - (\partial C(t) / \partial t )/ (2 C(t))$.  Using
\Eq(Langevin-1time-wn), this becomes
$\eta_C(t) = \eta(t) - \Re \eta(t) + D(t)/C(t)$.
It is interesting to note that this ensures $\Re \eta_C \ge 0$ even if $\Re
\eta < 0$.  These equations can be rearranged to give $\Im \eta(t) = \Im
\eta_C(t)$, $D(t) = C(t) \Re \eta_C(t)$, and $\Re \eta(t) = \Re \eta_C(t) -
(\partial \log C(t) / \partial t )/ 2$.  Thus, given any 3 arbitrary
functions $C(t) \ge 0$, $\Re \eta_C(t)$, and $\Im \eta_C(t)$ that determine
the model \Eq(Langevin-Cmod), it is possible to find a white-noise Langevin
equation for which it is the exact solution (as long as $\Re \eta_C \ge 0$
so that $D \ge 0$).  Conversely, for any arbitrary complex $\eta(t)$ and
real $D(t) \ge 0$ that specify a white-noise Langevin problem, one can
find a corresponding solution of the form \Eq(Langevin-Cmod).

It is interesting to note that
%
%
$C(t,t') =$ $C(t') \exp [ - \int_{t'}^t dt''\, \eta(t'')]$ (for
$t>t'$) is also an exact solution for this same white-noise Langevin
problem.  This form is valid for arbitrary $\eta(t)$ (even $\Re \eta < 0$).
However, BKO\cite{Bowman93} and references therein\cite{BKO-realiz}
indicate that this fails to preserve realizability when used in the context
of Markovian approximations to the DIA, so they instead use
\Eq(Langevin-Cmod).

On a related topic, BKO\cite{Bowman93} showed that their realizable
Markovian closure (RMC), as given by their Eqs.~(66a-e), has an underlying
Langevin representation given by their Eq.~(67) with a two-time noise
correlation function $\langle f_k(t) f_k^*(t') \rangle$ of the form of their
Eq.~(64), which is not necessarily white noise.  However, other two-time
noise correlation functions can also give the same equal-time statistics
equivalent to their Eq.~(66a).  This requires $F_k(t) = \Re \int_0^t d
\bar{t}\, \langle f_k(t) f_k^*(\bar{t}) \rangle R_k^*(t,\bar{t})$, where
$F_k(t)$ is the noise term in their Eq.~(66a).  For a case where $F_k(t)$
is always positive, then the RMC is also equivalent to a Langevin
representation with white-noise, $\langle f_k(t) f_k^*(t') \rangle = 2
D_f(t) \delta(t-t')$, where $D_f(t)$ $ = \half \Re \sum_{\vk + \vp + \vq
= 0} |M_{\vk \vp \vq}|^2 \Theta_{\vk \vp \vq} C_p^{1/2} C_q^{1/2}$.
While both 
white and non-white noise can give the same equal-time equations for
$C(t)$, they will give different results for the two-time correlation
function $C(t,t')$.  However, there can be cases where $F_\vk(t) < 0$, for
which a realizable Langevin representation must use non-white noise, as in
their Eq.~(64).  [Note that while $C(t) \ge 0$ is a fundamental requirement
preserved by a realizable theory, the ``triad interaction time'' $\Re
\Thkpq$ may go negative.  An example, similar to Eq.~(47) of
BKO,\cite{Bowman93} can be constructed for the realizable $\Thkpq$ of
Eq.~(66d) of BKO in the limit of constant $C_\vp$ and $C_\vq$ with
$\eta_\vk = \eta_\vp + \eta_\vq = \rho + i a$.]

\section{Fitting models to the two-time correlation function}
\label{Appendix-fit}

Conceptually the process of fitting an exponential model of
decorrelation to the actual two-time correlation function seems
straightforward.  But as described in Sec.~(\ref{Sec-Langevin-ss}) and
Sec.~(\ref{sec-Markov-ss}), there are various choices one could make in
the weights used to fit the models.  Galilean invariance imposes some
constraints, but does not completely constrain the problem.  In this
appendix we further describe some options and our choices.

Consider the following measure of the error between the actual two-time
correlation function and a model correlation function:
\begin{equation}
S(t) = \int_0^t dt'\, | C(t,t') - C_{\mod}(t,t') |^2 .
\label{S-err}
\end{equation}
We will assume $C_{\mod}(t,'t)$ is of the form of \Eq(Langevin-Cmod).  The
equal time correlation function $C(t)=C(t,t)$ is already specified, so
our task is to choose $\eta_C(t)$ in \Eq(Langevin-Cmod) in such a way as to
minimize the squared error $S$.  We want to stay in a Markovian
framework, where $\eta_C(t)$ depends on parameters only from the present
time.  We assume that $\eta_C(t')$ for times $t'<t$ has already been
chosen optimally.  But we can choose $\eta_C(t)$ at the present time so
that the extrapolation of $S(t)$ into the future is minimized.  That is, we
want to minimize $\partial S / \partial t$, which, after using
\Eq(Langevin-2time) for $\partial C(t,t') / \partial t$ and
\Eq(Langevin-Cmod) to evaluate $\partial C_{\mod}(t,t') / \partial t$, is 
%
\begin{eqnarray}
{\partial S \over \partial t} & = &
 2 \int_0^t dt' \,
   \left[ - \eta(t) C(t,t') + \int_0^{t'} d \bar{t}\, R^*(t',\bar{t})
   C_f^*(t,\bar{t}) \right. \nonumber \\
&& \left. - {1 \over 2 C(t)} {\partial C(t) \over \partial t} C_{\mod}(t,t')
   + \eta_C(t) C_{\mod}(t,t') \right] \nonumber \\
&&\quad \times (C^*(t,t') - C_{\mod}^*(t,t'))
+ {\rm c.c.},
\label{dSdt-min}
\end{eqnarray}
where ${\rm c.c.}$ indicates the complex conjugate of the previous
expression.  
Separately minimizing $\partial S / \partial t$ with respect to the real
part $\eta_{Cr}$ and imaginary part $\eta_{Ci}$ of $\eta_C(t)$
(i.e., set $\partial (\partial S / \partial t) / \partial
\eta_{Cr} = 0$, and then $\partial (\partial S / \partial t) / \partial
\eta_{Ci} =0$) leads to the requirement that
\begin{equation}
\int_0^t dt'\, C_{\mod}^*(t,t') C_{\mod}(t,t')
= \int_0^t dt'\, C_{\mod}^*(t,t') C(t,t').
\label{A-best-fit}
\end{equation}
[Note that when evaluating derivatives of \Eq(dSdt-min) with respect to
$\eta_{Cr}$ and $\eta_{Ci}$, it is only the explicit appearance of
$\eta_C(t)$ in \Eq(dSdt-min) that is important.  The parameter
$\eta_C(t)$ also appears implicitly via the definition of
$C_{\mod}(t,t')$, but there it has an impact on the integral defining
$\partial S/\partial t$ only through a set of measure zero, and so can
be neglected as long as $\eta_C(t)$ is bounded.]  

In the steady-state limit, \Eq(A-best-fit) is equivalent to
\Eq(etac-invariant).
For a time-dependent case, consider \Eq(A-best-fit) as providing
a constraint of the form $A_{\mod}(t) = A(t)$.  Assuming that this has
already been satisfied for earlier times, we want it to remain
satisfied for future times, i.e., we need to require that $\partial
A_{\mod} / \partial t = \partial A / \partial t$.  This is precisely what
we are doing when we set \Eq(Langevin-Adot1) and \Eq(Langevin-Adot2) to
be equal, and it leads to a formula for $\eta_C(t)$ at the present time
that minimizes the errors as time advances.

The same procedures as described here are used in fitting a model
response function $R_{\mod}(t,t')$ of the form of \Eq(Rmodk) to the
actual response function, leading to the constraint
\begin{equation}
\int_0^t dt'\, R_{\mod}^*(t,t') R_{\mod}(t,t')
= \int_0^t dt'\, R_{\mod}^*(t,t') R(t,t')
\label{R-best-fit}
\end{equation}
As mentioned at the end of Sec.~(\ref{sec-Markov-ss}), $R_{\mod}^*$
in this expression could be replaced with $C_{\mod}^*$ and one would
still get an expression defining $\eta$ that was Galilean invariant.
However, \Eq(R-best-fit) seems to make more sense as a least-squares
best fit of $R_{\mod}$ to $R$, and that is the choice we have
made.

But consider \Eq(S-err) in the steady-state limit where $\eta_C$ is a
constant and $C(t,t')$ depends only on $t-t'$,
\begin{equation}
S_0 = \int_{-\infinity}^t dt'\, | C(t,t') - C_0 e^{-\eta_C (t-t')} |^2 .
\label{S-err-ss}
\end{equation}
It is straightforward to show that choosing $\eta_C$ to minimize the
total squared error $S_0$ leads to the condition
\begin{eqnarray}
&& \int_{-\infinity}^t dt'\, e^{-\eta_C^*(t-t') } C_0 e^{-\eta_C (t-t')}
(t-t') \nonumber \\
& = & \int_0^t dt'\, e^{-\eta_C^*(t-t')} C(t,t') (t-t').
\label{At-best-fit}
\end{eqnarray}
Note that this differs from \Eq(A-best-fit) by an additional factor of
$(t-t')$, which weights errors at larger time
separation more strongly.  Including an extra weighting factor of $(t-t')$ in
\Eq(etac-invariant) might help to refine the model, particularly for cases
such as in Fig.~(5), where the short time behavior is reasonable but
the long-time fit needs improvement.
%

It is perhaps not surprising that optimizing a constant $\eta_C$ to
minimize the global error $S_0$ gives a somewhat different result than
optimizing $\eta_C(t)$ to minimize the local error $\partial S /
\partial t$.  
In order for the time-dependent fitting procedures to reproduce this
steady-state result, one could modify \Eq(S-err) by multiplying the
integrand by a factor of $(t-t')$.  Working through the derivation, one
finds that the integrands in \Eq(A-best-fit) would be modified to also
have an additional weighting factor of $(t-t')$.  Thus one might be able
to improve the results in this paper some by including an extra
weighting of $(t-t')$ in the appropriate places, \Eq(Langevin-Area),
\Eq(Rk-ss-int), \Eq(Ak-int), and \Eq(Ack-int), and working through the
derivations to see the modified results.  While such modifications could
lead to an improved model, and would be interesting for future work, one
should realize that the dynamics are complicated and no choice of
weights is perfect.  For example, what one really wants is a best fit
model for the triad interaction times which are weighted by interactions
between three modes as given in \Eq(DIACeqBTh d), not necessarily best
fits for the decorrelation rates of just individual modes.  Probably a
higher priority for future work is to use a starting set of equations
that satisfy random Galilean invariance, so that interactions with large
scales are not overemphasized as they are in the Eulerian DIA.

\section{Inertial-range scaling of DIA-based closures}
\label{Appendix-scaling}
Here we determine steady-state self-similar inertial-range
solutions in $d$ dimensions to closures of the form~\hide\Eq(DIACeqBTh)
in an unbounded domain (so that $\SD\goesto
\int_{\D_\vk} d\vp\,d\vq \doteq \int d\vp\,d\vq\,\d(\vk+\vp+\vq)$),
taking the initial time $t_0=-\infty$.
This extends previous derivations in the literature to self-similar
spectra consistent with generic DIA-based closures~\hide\Eq(DIACeqBTh) of
the quadratically nonlinear equation~\hide\Eq(master),
arising from the cascade of a generalized energy $\half\sum_\vk\sigk
\Abs{\psik(t)}^2$. Assuming self-similar scalings of the mode-coupling
and statistical variables, our derivation requires only the additional
condition~\hide\Eq(steadystateRint), which is somewhat weaker than statistical
stationarity.

The turbulence could be forced with a
linear instability, incorporated with dissipation into the linear
coefficient $\nuk$, or else a random force could be added to the right-hand
side of \Eq(DIACeqBTh a).
By definition, both the external forcing and dissipation~$\nuk$ vanish in the
inertial range. The symmetry~\hide\Eq(Menergysym) then implies that the
nonlinear terms in \Eq(DIACeqBTh a), weighted by $\s_\vk$, must balance. It
is convenient to define
\begin{eqnarray*}
&&S_\vk\doteq\s_\vk\Re(\Fk-\Nk)\nonumber\\
&=&\half\Re\int_{\D_\vk} d\vp\,d\vq\,\s_\vk\Mkpq\Mkpq^*\BThkpq\nonumber\\
&&\quad+\Re\int_{\D_\vk} d\vp\,d\vq\,\s_\vk\Mkpq\Mpqk^*\BThpqk^*\nonumber\\
&=&-\half\Re\int_{\D_\vk} d\vp\,d\vq\,\Mkpq(\s_\vp\Mpqk^*+\s_\vq\Mqkp^*)\BThkpq\nonumber\\
&&\quad+\Re\int_{\D_\vk} d\vp\,d\vq\,\s_\vk\Mkpq\Mpqk^*\BThpqk^*\nonumber\\
&=&-\Re\int_{\D_\vk} d\vp\,d\vq\,\s_\vp\Mkpq\Mpqk^*\BThkpq\nonumber\\
&&\quad+\Re\int_{\D_\vk} d\vp\,d\vq\,\s_\vk\Mkpq\Mpqk^*\BThpqk^*\nonumber\\
&=&\Re\int_{\D_\vk} d\vp\,d\vq\,\Mkpq\Mpqk^*(\s_\vk\BThpqk^*-\s_\vp\BThkpq).
\end{eqnarray*}

We seek self-similar solutions of the DIA that obey the scalings (for $\l>0$)
\begin{mathletters}\eq(scaling)
\begin{equation}
M_{\l\vk,\l\vp,\l\vq}=\l^m M_{\vk\vp\vq},
\end{equation}
\begin{equation}
\s_{\l\vk}=\l^s \s_{\vk},
\end{equation}
\begin{equation}
R_{\l\vk}(t,t')=R_{\vk}(t,t-\lambda^{-\ell}(t-t')),
\end{equation}
\begin{equation}
C_{\l\vk}(t,t')=\l^n C_{\vk}(t,t-\lambda^{-\ell}(t-t')),
\end{equation}
so that, upon making the change of variables
$\bar s\doteq t-\lambda^{-\ell}(t-\tb)$ in \Eq(DIACeqBTh d),
\begin{equation}
\BTh_{\l\vk,\l\vp,\l\vq}=\l^{\ell+2n} \BTh_{\vk\vp\vq}.
\end{equation}
\end{mathletters}
Once we have determined suitable values of the scaling exponent $n$,
we may compute the wavenumber exponent
$\b$ for the energy spectrum $E(k)\sim \e^\a k^\b$. If the total energy $E$
is related to the correlation function $\Ck$ of the fundamental variable
$\psi$ by $E=\int d\vk\,k^\g\Ck=\int dk \, E(k)$, then $\b=d-1+\g+n$.

Following Ref~\onlinecite{Orszag73}, we will use the change of variables
$z=k^2/p$, $w=kq/p$
to determine values of the exponents $\ell$ and $n$ for which the angular average $S(k)$ of $S_\vk$ vanishes.
In terms of the scaling factor $\l=k/z$ we note that
$k=\l z$, $p=\l k$, and $q=\l w$.
Letting $\vz=z\phat$ and $\vw=w\qhat$, we may then express
$d\vp\,d\vq=\l^{3d} d\vz\,d\vw$ and
$\d(\vk+\vp+\vq)=\l^{-d}\d(z\khat+k\phat+\vw)$.
Hence, upon interchanging $\phat$ and $\khat$ in the integration, we deduce
\begin{eqnarray*}
S(k)&\doteq&\int d\khat\,S_\vk=\Re\int d\khat
\int_{\D_\vk} d\vz\,d\vw\,\l^{3d-d+2m+s+\ell+2n}\nonumber\\
&&\quad\times M_{\vz,\vk,\vw}M_{\vk,\vw,\vz}^*
(\s_\vz\BTh_{\vk,\vw,\vz}^*-\s_\vk\BTh_{\vz,\vk,\vw})\nonumber\\
&=&-\Re\int d\khat
\int_{\D_\vk} d\vz\,d\vw\,\l^{2d+2m+s+\ell+2n}\nonumber\\
&&\quad\times M_{\vk,\vz,\vw}^* M_{\vz,\vw,\vk}
(\s_\vk\BTh_{\vz,\vw,\vk}-\s_\vz\BTh_{\vk,\vz,\vw}^*)\nonumber\\
&=&-S(k),
\end{eqnarray*}
provided that
\begin{equation}
2d+2m+s+\ell+2n=0.\eq(Cscaling)
\end{equation}
The condition~\hide\Eq(Cscaling) guarantees that the angle-averaged
nonlinear terms in \Eq(DIACeqBTh a) will balance in a steady state
and lead to an inertial range.

The exponent $\ell$ can be determined by integrating the
DIA response function equation
\begin{eqnarray}
&&\delt\Rk(t,t')
- \I{-\infty}t d\tb\int_{\D_\vk} d\vp\,d\vq\,\Mkpq\Mpqk^*\nonumber\\
&&\quad\times 
\Rp^*(t,\tb)\,\Cq^*(t,\tb)\,\Rk(\tb,t')
	= \d(t-t'),\eq(DIAR)
\end{eqnarray}
over all $t'$, using the steady-state condition
\begin{equation}
\lim_{t\goesto\infty}\delt{}\I{-\infty}{\infty} dt'\,R(t,t')=0.
\eq(steadystateRint)
\end{equation}
One obtains
\begin{eqnarray}
&&-\I{-\infty}{\infty} d\tb\int_{\D_\vk} d\vp\,d\vq\,\Mkpq\Mpqk^*\nonumber\\
&&\quad\times
\Rp^*(\infty,\tb)\,\Cq^*(\infty,\tb)\I{-\infty}{\infty} dt'\,\Rk(\tb,t')=1.\eq(Rbalance)
\end{eqnarray}
Upon replacing $\vk$ by $\lambda\vk$ (for any constant $\lambda$) and
exploiting the self-similar scalings given in \Eqs(scaling), we make
the change of variable $s'=\tb-\lambda^{-\ell}(\tb-t')$ to obtain
\begin{eqnarray*}
&&-\l^{d+2m+\ell+n}\I{-\infty}{\infty} d\tb\int_{\D_\vk}
d\vp\,d\vq\,\Mkpq\Mpqk^*\nonumber\\
&&\quad\times\Rp^*(\infty,\bar s)\,\Cq^*(\infty,\bar s)
\I{-\infty}{\infty}ds'\,\Rk(\tb,s')=1,
\end{eqnarray*}
where $\bar s\doteq t-\lambda^{-\ell}(t-\tb)$.
The integral over $\tb$ is dominated by contributions from large
$\tb$, for which the integral over $s'$
asymptotically approaches a constant (with respect to~$\tb$),
according to \Eq(steadystateRint). Hence, after making a final change of
variables from $\tb$ to $\bar s$, we see that the balance expressed in
\Eq(Rbalance) is recovered if
\begin{equation}
\l^{d+2m+2\ell+n}=1,\eq(Diverge)
\end{equation}
from which we conclude that $\ell=-(d+n)/2-m$. If one inserts this result
into \Eq(Cscaling), one obtains the Kolmogorov scalings
\begin{mathletters}\eq(Kolmogorovscalings) 
\begin{equation}
\ell=\fr{1}{3}s-\fr{2}{3}m,
\end{equation}
\begin{equation}
n=-d-\fr{2}{3}(m+s),\eq(Knscaling)
\end{equation}
\begin{equation}
\b=\g-1-\fr{2}{3}(m+s).
\end{equation}
\end{mathletters}
Alternatively, one could adopt instead of \Eq(steadystateRint) the
stronger condition of \It{statistical stationarity},
$\Rk(t,t')={\cal R}_\vk(t-t')$ and $\Ck(t,t')={\cal
C}_\vk(t-t')$. \Equation(Diverge) is 
then readily seen to follow directly from \Eq(DIAR).
In either case we have only shown that \Eq(Diverge) is a necessary
condition for self-similar solutions of the form~\hide\Eq(scaling) to exist. 
In order that these solutions actually satisfy \Eq(DIAR), it is also
necessary at the very least that the wavenumber integral in \Eq(DIAR)
converges.

Unfortunately, the scaling expressed in \Eq(Diverge) often leads to a
divergence of the $q$ integral in \Eq(DIAR), preventing self-similar
solutions from existing. Typically, the mode-coupling coefficients
$M_{\vk,-\vk-\vq,\vq}$ asymptotically approach a constant as $q$ goes to
zero while $\vk$ is held fixed. Upon performing the $\vp$ integration in
\Eq(DIAR), we then see that the $q$ integrand will scale like
$q^{d-1}\Cq^*(t,\bar t)$ for small $q$. If $\Cq$ asymptotically scales as
$q^n$, then the integrand will scale like $q^{d-1+n}$.  But \Eq(Knscaling)
implies that $d-1+n=-1-2(m+s)/3$. Normally $m+s > 0$ (see
Table~\ref{cascades}); in these cases there would be a divergence of the $q$
integral in \Eq(DIAR) if self-similar solutions really were to exist.
\cite{Edwards64,Leslie73b}

This divergence indicates that the dominant contributions to the
eddy-turnover time come from the energy spectrum at large scales, where
self-similarity no longer holds.  (For this reason, the DIA is not
invariant to random Galilean transformations.)  The actual value of the
scaling $\ell$ that appears in the DIA response must be calculated by
taking into account that $\Cq$ does not actually behave as $q^n$ for small
$q$. The DIA equations apply to the case of zero mean flow, where the
energy spectrum goes to zero at small wavenumbers. This means that the
integration in \Eq(DIAR) must be effectively cut off at some fixed large
scale wavenumber $k_0$. The introduction of this cutoff wavenumber removes
the divergence in the integral, but it also changes the above scaling
argument. Since the dominant contribution to \Eq(DIAR) still comes from
small $q$, we need to identify the scaling of the mode-coupling
coefficients with $k$ for $q\muchl k$, $M_{\l\vk,-\l\vk,\l\vq}=\l^{m'}
M_{\vk,-\vk,\vq}.$ Since the lower wavenumber limit is
now fixed, no self-similar scaling in $\vq$ can be made; the scaling with
$k$ for small $q$ then leads to $\l^{2\ell+2m'}=1$.  
Hence for the DIA equations the actual scalings of the response function,
correlation function, and energy spectrum are given by
\begin{mathletters}\eq(Kraichnanscalings)
\begin{equation}
\ell_{\rm DIA}=-m',
\end{equation}
\begin{equation}
n_{\rm DIA}=-d-m+\fr{m'-s}{2},
\end{equation}
\begin{equation}
\b_{DIA}=\g-1-m+\fr{m'-s}{2}.
\end{equation}
\end{mathletters}

In Table~\ref{cascades} we compare the scalings in \Eqs(Kolmogorovscalings)
with the anomalous DIA scalings given by \Eq(Kraichnanscalings). 
The scalings given by \Eq(Kolmogorovscalings) are consistent with
Kolmogorov's dimensional analysis. We emphasize that these scalings would have
also been obtained for the DIA equations (they too are dimensionally
consistent) had the wavenumber integral in \Eq(Rbalance) converged.

\begin{table}
\begin{tabular}{|l|c|c|c|c|c|c||c|c|c||c|c|c|}
	Cascade&$\psi$&$d$&$s$&$\g$&$m$&$m$'&
$\ell$&$n$&$\b$&$\ell_{\rm DIA}$&
$n_{\rm DIA}$&$\b_{\rm DIA}$\\
\hline
&&&&&&&&&&&&\\
	2D enstrophy&$\Psi$&$2$&$4$&$2$&$2$&$1$&$0$&$-6$&$-3$&
$-1$&$-\fr{11}{2}$&$-\fr{5}{2}$\\
&&&&&&&&&&&&\\
	2D energy&$\Psi$&$2$&$2$&$2$&$2$&$1$&$-\fr{2}{3}$&$-\fr{14}{3}$&$-\fr{5}{3}$&
$-1$&$-\fr{9}{2}$&$-\fr{3}{2}$\\
&&&&&&&&&&&&\\
&&&&&&&&&&&&\\
	3D energy&$u$&$3$&$0$&$0$&$1$&$1$&$-\fr{2}{3}$&$-\fr{11}{3}$&$-\fr{5}{3}$&
$-1$&$-\fr{7}{2}$&$-\fr{3}{2}$\\
&&&&&&&&&&&&\\
&&&&&&&&&&&&\\
	3D helicity&$u$&$3$&$1$&$0$&$1$&$1$&$-1$&$-\fr{13}{3}$&$-\fr{7}{3}$&
$-1$&$-4$&$-2$\\
&&&&&&&&&&&&\\
\end{tabular}
\caption{Scaling exponents for various cascades in two dimensions (2D) and
three dimensions (3D), using either the streamfunction $\psi=\Psi$ or
velocity $\psi=u$ normalization.}\label{cascades}  
\end{table}

\bibliographystyle{pf}
\bibliography{refs,mrm-foot}


\end{document}